\documentclass[sn-mathphys]{sn-jnl}
\jyear{2022}%

\usepackage{graphicx}

\newcommand{\hrho}{\hat{\rho}}
\newcommand{\hsigma}{\hat{\sigma}}
\newcommand{\ha}{\hat{a}}

\newcommand{\bbP}{\mathbb{P}}

\newcommand{\vb}{\mathbf{b}}
\newcommand{\vc}{\mathbf{c}}

\newcommand{\vt}{\mathbf{t}}

\newcommand{\vv}{\mathbf{v}}

\newcommand{\vx}{\mathbf{x}}
\newcommand{\vy}{\mathbf{y}}
\newcommand{\vz}{\mathbf{z}}
\newcommand{\valpha}{\boldsymbol{\alpha}}

\newcommand{\mA}{\mathbf{A}}
\newcommand{\mB}{\mathbf{B}}
\newcommand{\mC}{\mathbf{C}}
\newcommand{\mD}{\mathbf{D}}

\newcommand{\mI}{\mathbf{I}}

\newcommand{\mN}{\mathbf{N}}

\newcommand{\mS}{\mathbf{S}}

\newcommand{\mV}{\mathbf{V}}

\newcommand{\mZ}{\mathbf{Z}}

\newcommand{\mOmega}{\boldsymbol{\Omega}}

\newcommand{\mSigma}{\boldsymbol{\Sigma}}

\newcommand{\Expect}{\mathbb{E}}
\newcommand{\intd}{\,\mathrm{d}}

\newcommand{\tr}{\mathrm{tr}}
\newcommand{\Var}{\mathrm{Var}}

\usepackage{amsmath}
\DeclareMathOperator*{\argmax}{arg\,max}

\usepackage{accents}

\usepackage{amsmath,amsfonts,amssymb}
\usepackage{caption}
\usepackage{subcaption}
\usepackage{bm}
\usepackage{multirow}

\usepackage{mathrsfs}
\usepackage{amsbsy}
\usepackage{titletoc}
\usepackage{color}
\usepackage{comment}

\def\cov{{\mbox{cov}}}
\def\tr{{\mbox{tr}}}

\def\b{{\mathbf{b}}}

\def\z{{\mathbf{z}}}
\def\btheta{{\boldsymbol{\theta}}}

\def\v1{{\mathbf{1}}}
\theoremstyle{thmstyleone}%
\newtheorem{theorem}{Theorem}[section]
\newtheorem{assumption}{Condition}
\newtheorem{lemma}[theorem]{Lemma}

\theoremstyle{thmstyletwo}%

\theoremstyle{thmstylethree}%

\usepackage[algo2e,ruled,vlined,linesnumbered]{algorithm2e}

\normalbaroutside 

\begin{document}
\title[Batch effect correction with sample remeasurement]{Batch effect correction with sample remeasurement in highly confounded case-control studies}


\author[1]{\fnm{Hanxuan} \sur{Ye}}\email{hanxuan@tamu.edu}

\author*[1]{\fnm{Xianyang} \sur{Zhang}}\email{zhangxiany@stat.tamu.edu}
\author[2]{\fnm{Chen} \sur{Wang}}\email{wang.chen@mayo.edu}
\author[2]{\fnm{Ellen L.} \sur{Goode}}\email{egoode@mayo.edu}
\author*[2]{\fnm{Jun} \sur{Chen}}\email{chen.jun2@mayo.edu}

\affil[1]{\orgdiv{Department of Statistics}, \orgname{Texas A\&M University}, \orgaddress{\street{155 Ireland Street}, \city{College Station}, \postcode{77843}, \state{TX}, \country{USA}}}

\affil[2]{\orgdiv{Department of Quantitative Health Sciences}, \orgname{Mayo Clinic}, \orgaddress{\street{200 First Street SW}, \city{Rochester}, \postcode{55905}, \state{MN}, \country{USA}}}


\abstract{
Batch effects are pervasive in biomedical studies. One approach to address the batch effects is repeatedly measuring a subset of samples in each batch. These remeasured samples are used to estimate and correct the batch effects. However, rigorous statistical methods for batch effect correction with remeasured samples are severely underdeveloped. In this study, we developed a framework for batch effect correction using remeasured samples in highly confounded case-control studies. We provided theoretical analyses of the proposed procedure, evaluated its power characteristics, and provided a power calculation tool to aid in the study design. We found that the number of samples that need to be remeasured depends strongly on the between-batch correlation. When the correlation is high, remeasuring a small subset of samples is possible to rescue most of the power. 
}

\maketitle

\section{Introduction}\label{sec:Intro}
One major issue facing biological studies is that biological measurement is highly susceptible to non-biological experimental variation or ``batch effects". Batch effects are pervasive in modern high-throughput omics technologies using microarrays or next-generation sequencing~\citep{leek2010tackling, goh2017batch}. Different experimental conditions, measurement modalities, personnel executing the experiments, and batches of reagents all contribute to batch effects. Such unwanted variation has severe statistical consequences. It could reduce statistical power by introducing extra variation or, more seriously, lead to false findings if the batch effects are confounded with the effects of interest. Although performing the biological measurement in a single batch is the most effective way to reduce batch effects, such practice may not always be possible due to various constraints such as resource availability and measuring capacity.   Even if the experimental measurement is executed in a single batch, some unexpected batch effects could still arise. For example, different measuring chips, locations on the chips,  DNA extraction plates, and sequencing lanes have all been found to produce batch effects in omics studies~\citep{scherer2009batch,tom2017identifying,price2018adjusting}. Therefore, 
addressing the batch effects in the study design and data analysis is critical to improve the statistical power, increase the robustness of the findings and reduce the developmental cost.

Over the past two decades, a number of batch effect correction methods have been developed and applied in practical data analysis.
Two mainstreams for batch effect correction are location-scale (LS) matching and matrix factorization (MF). The LS methods assume the sources of the batch effects are known so that the location (e.g., mean), scale (e.g., standard deviation), or even the entire distribution are matched across batches. Methods in this category include batch mean-centering (BMC)~\citep{sims2008removal}, gene-wise standardization (SD)~\citep{li2001model}, ComBat~\citep{johnson2007adjusting, 10.1093/nargab/lqaa078}, cross-platform normalization (XPN)~\citep{shabalin2008merging} and distance-weighted discrimination (DWD)~\citep{benito2004adjustment}. Among these, ComBat, an empirical Bayes-based LS method, is the most widely used method due to its robustness to small batch sizes compared with earlier methods ~\citep{johnson2007adjusting, 10.1093/nargab/lqaa078}. In contrast, the MF-based methods do not require the sources of batch effects are known in advance. Instead, they search for directions of maximal variance associated with the batch effects and use the resulting latent factors to correct for batch effects. Methods in this category include singular value decomposition (SVD)/principal component analysis (PCA) \citep{alter2000singular,jolliffe2013principal}, surrogate variable analysis (SVA)~\citep{leek2007capturing}, {  RUV~\citep{gagnon2012using, gagnon2013removing, jacob2016correcting} and LEAPP \citep{sun2012multiple}.  The SVA, RUV, and LEAPP have also been studied and expanded within a unified CATE rotation \citep{wang2017confounder} framework that adjusts for the confounders in hypothesis testing. 
}

Previous efforts for batch effects correction have been focused on estimating and correcting batch effects based on independent samples~\citep{leek2007capturing, johnson2007adjusting, 10.1093/nargab/lqaa078,sun2012multiple,wang2017confounder}. In practice, however, one intuitive approach used by investigators to address batch effects is through remeasuring a subset of samples in each batch in the hope that these remeasured samples could be used to estimate and correct the batch effects~\citep{tasaki2018multi, xia2021batch}. Unfortunately, other than some simple approaches, statistical methods for batch correction using the remeasured samples remain severely underdeveloped.  Biostatisticians are often faced with the inability to efficiently utilize these remeasured samples in the analysis to correct for batch effects,  hindering the successful completion of the proposed studies. To fill the methodological gap, this study investigates the feasibility and methodology for batch effect correction using remeasured samples in a highly confounded case-control study~\citep{zhou2019examining}. We specifically consider a challenging scenario, where  an investigator has collected all the case samples, and she wants to compare these case samples to the control samples that have already been measured previously { and a subset of which are still available for  remeasurement}. This scenario is quite common in clinical settings since clinical investigators usually obtain case samples more easily than control samples. {For example, an investigator wants to compare her case samples to the control samples from the institutional biobank \citep{olson2019characteristics}.  Oftentimes, the biobank samples have already been characterized in a standalone study or have been used as controls in other disease studies, resulting in a large amount of pre-existing control data that can be potentially used together with the new control data generated from remeasuring a subset of the biobank samples. Another example is the subsequent analysis in case-cohort studies \citep{rundle2005design}. One  strength of case-cohort studies is that the subcohort can be used as a reference group for a variety of different case groups. The subcohort in a case-cohort study implemented early for a common disease can be used  as a reference group for a series of rarer or long-latency disease. The data for the reference group already exists in study databases. For subsequent disease studies, the existing subcohort data may be re-used after a subset of the subcohort samples have been remeasured. }  

Obviously, if none of the control samples are remeasured,  the biological effects will be completely confounded with the batch effects, and distinguishing between the biological and batch effects will be very difficult.  Ideally, all the control samples need to be remeasured together with the case samples to maximize the discovery power. 
However,  due to resource constraints and sample availability, such practice may not always be possible. Therefore, it will be of tremendous help if only remeasuring a small subset of control samples is necessary to correct the batch effects. 
Despite of a subject of critical importance, to our surprise, no dedicated statistical methods are available. No theoretical investigation has been performed to study the operating characteristics of batch effect correction with remeasured samples. It is unknown how many control samples need to be remeasured to recover most of the power, whether a handful of controls are sufficient to correct for batch effects, and what factors matter most in deciding the number of remeasured samples. A rigorous statistical testing method coupled with a power calculation tool is critically needed for this particular scenario. A successful tool could potentially rescue a completely confounded study and has a tremendous economic impact on the field.

In this study, we proposed a computationally efficient statistical method for batch effect correction with remeasured samples for a highly confounded case-control study. The method is based on the maximum likelihood framework, and hence the derived procedure is optimal in using the information available. We studied the theoretical properties of the procedure and proved the consistency and asymptotic normality of the resulting estimators. 
We investigated the power characteristics of the approach based on simulations and theoretical analysis, and identified statistical properties affecting its power. Finally, we proposed a  power calculation tool to aid in the study design. A real dataset with known batch effects and a large number of remeasured samples was used to demonstrate the feasibility and efficiency of the proposed procedure.

\section{Results}

\begin{figure}[t!]
    \centering
    \includegraphics[width = 0.8\textwidth]{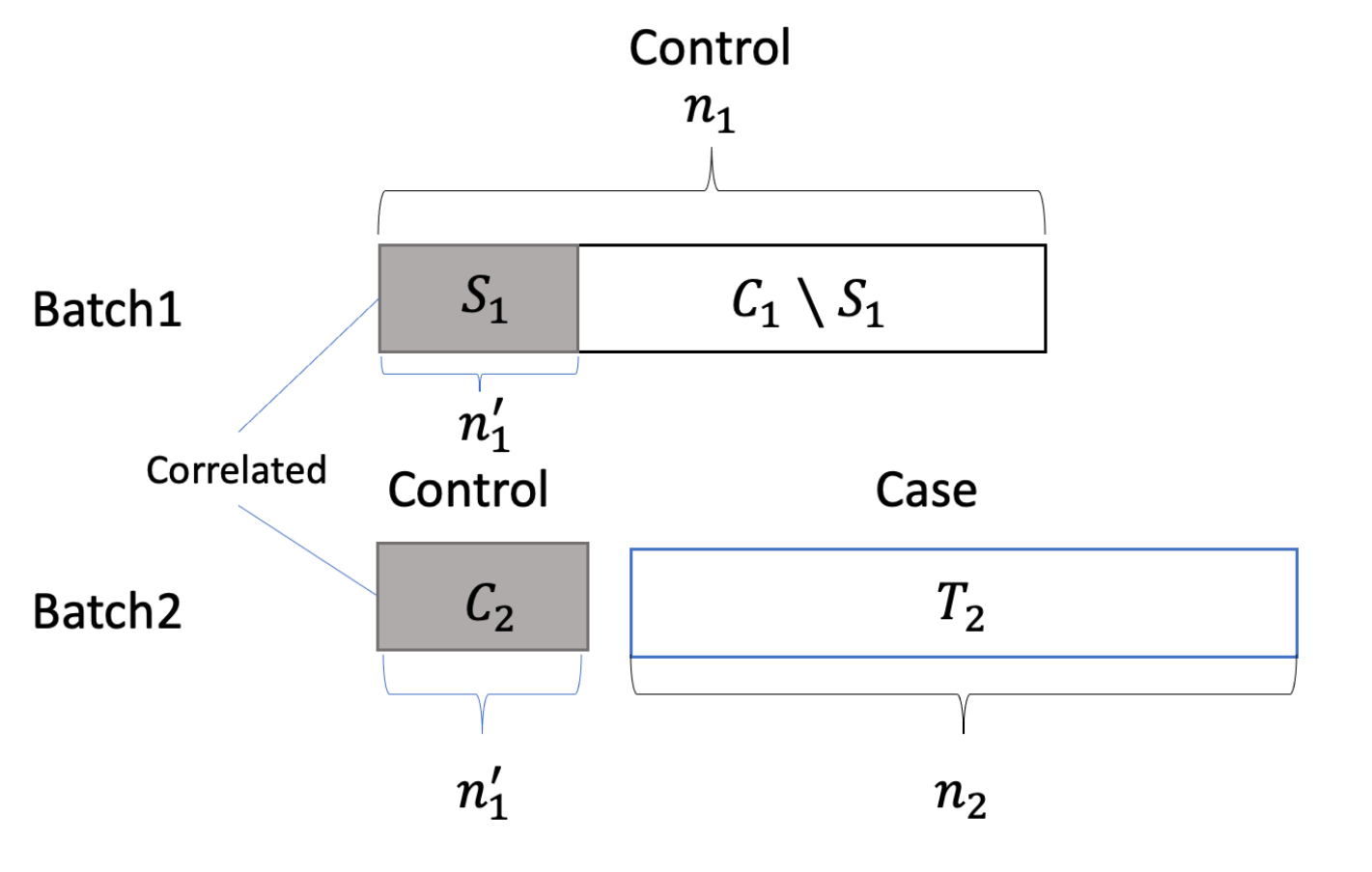}
   \caption{ {\bf Illustration of the study design.} Here, $C_1$ denotes the set of $n_1$ control samples in Batch 1, and $T_2$ denotes the set of case samples in Batch 2. A subset $S_1$ consisting of $n_1'$ samples  from the $C_1$ set is remeasured. The set of these remeasured samples in Batch 2 is indicated as $C_2$. $ C_1\setminus S_1$ represents the unmeasured control samples.}
    \label{fig:model}
\end{figure}
\addcontentsline{mtf}{figure}{\numberline{\thefigure}{\bf Illustration of the study design.} Here, $C_1$ denotes the set of $n_1$ control samples in Batch 1, and $T_2$ denotes the set of case samples in Batch 2. A subset $S_1$ consisting of $n_1'$ samples  from the $C_1$ set is remeasured. The set of these remeasured samples in Batch 2 is indicated as $C_2$. $ C_1\setminus S_1$ represents the unmeasured control samples. }  

\subsection{Problem Setup and Model}\label{subsec:model}

Consider that the control and case samples are measured on two different batches. We assume the linear model 
\begin{equation}\label{eq1} 
\begin{split}
&y_i =  x_i(a_{0} + a_{1}) +  \z_i^\top \b + \epsilon_i, 
\\ & \epsilon_{i} | x_i \sim N(0, (1-x_i)\sigma_{1}^2 +  x_i \sigma_{2}^2), 
\end{split}
\end{equation} 
for $i=1,2,\dots,n$, where $y_i$ is the outcome, $x_i \in \{0, 1\}$ is the control/case group membership ($0$: control, $1$: case), $\z_i$ contains measurements of other covariates  including the intercept and possible covariate-batch (group) interactions, $a_{0}$ is coefficient for the true biological effect and $a_{1}$ is the coefficient for the nuisance batch effects. Since the batch effect and biological effect are indistinguishable in this example,  remeasurement of a subset of samples is thus necessary. Suppose the control and case samples are collected in the first and second batch, respectively, and a subset of control samples of size $n'$ are remeasured in the second batch. Suppose the control and case group contain $n_1$ and $n_2$ samples, respectively.
Without loss of generality, we assume that the first $n_1'$ control samples are remeasured, where $n_1'\leq n_1$ (Figure \ref{fig:model}).  Then we have
\begin{align*}
&\text{Control (batch 1):} \quad y_i = \z_i^\top \b + \epsilon_i^{(1)},\quad i=1,2,\dots,n_1,\\
&\text{Case (batch 2):} \quad y_i = a_{0}+a_{1} + \z_i^\top \b + \epsilon_i^{(2)},\quad i=n_1+1,\dots,n_1+n_2=n,\\
&\text{Control (batch 2):} \quad y_i = a_{1}+\z_i^\top \b + \epsilon_i^{(2)},\quad i=n+1,\dots,n+n_1',
\end{align*}
where $\epsilon_{i}^{(1)} \sim N(0, \sigma_{1}^2)$ for $1\leq i\leq n_1$, $\epsilon_{i}^{(2)} \sim N(0, \sigma_{2}^2)$ for $n_1+1\leq i\leq n+n_1'$, 
and $$\textrm{cov}(\epsilon_{i}^{(1)}, \epsilon_{n+i}^{(2)}) = \rho \sigma_{1} \sigma_{2}$$ 
for $1\leq i\leq n_1'$. The goal here is to develop an efficient procedure to test the null hypothesis that
$$H_0: a_{0}=0,$$
i.e., there is no true biological effect.

We introduce some notation before describing the estimation and inference procedures. Denote by $C_1$ the set of control samples in the first batch and $T_2$ the set of case samples in the second batch. Let $S_1=\{1,\dots,n_1'\}$ and $C_2=\{n+1,\dots,n+n_1'\}$ be the subset of remeasured control samples in batch 1 and batch 2, respectively, where $|S_1|= |C_2| = n_1'$. See Figure \ref{fig:model} for illustration. Note that the covariates associated with the samples in $S_1$ and $C_2$ are the same. 
Let $\btheta = (a_0, a_1, \vb, \rho, \sigma_1, \sigma_2)$ be the parameter vector to be estimated, and $\mN = (n_1, n_2, n_1')$ be the vector of sample sizes. We define $\mu_{1i} = \vz_i^\top \vb$ for $i \in C_1 $, $\mu_{2i} = a_0 + a_1 + \vz_i^\top \vb$ for $i \in T_2 $ and $\mu_{3i} = a_1 + \vz_i^\top \vb$ for $i \in C_2$.

\subsection{Simulation Studies}\label{sec:simulation}
We conduct a set of simulation studies to investigate the finite sample performance of the proposed procedure in terms of estimation accuracy, type I error rate, and statistical power. Moreover, we compare our method (``ReMeasure'') with three alternative procedures:
\begin{enumerate}
\item The location-scale matching approach (``LS", details in Supplementary Section 3).
\item Estimation and inference using only the second batch data (``Batch2").
\item Estimation and inference using the whole data set while ignoring the batch effects (``Ignore").
\end{enumerate}

\begin{figure}[ht!]
    \centering
    \includegraphics[width = 0.9\textwidth]{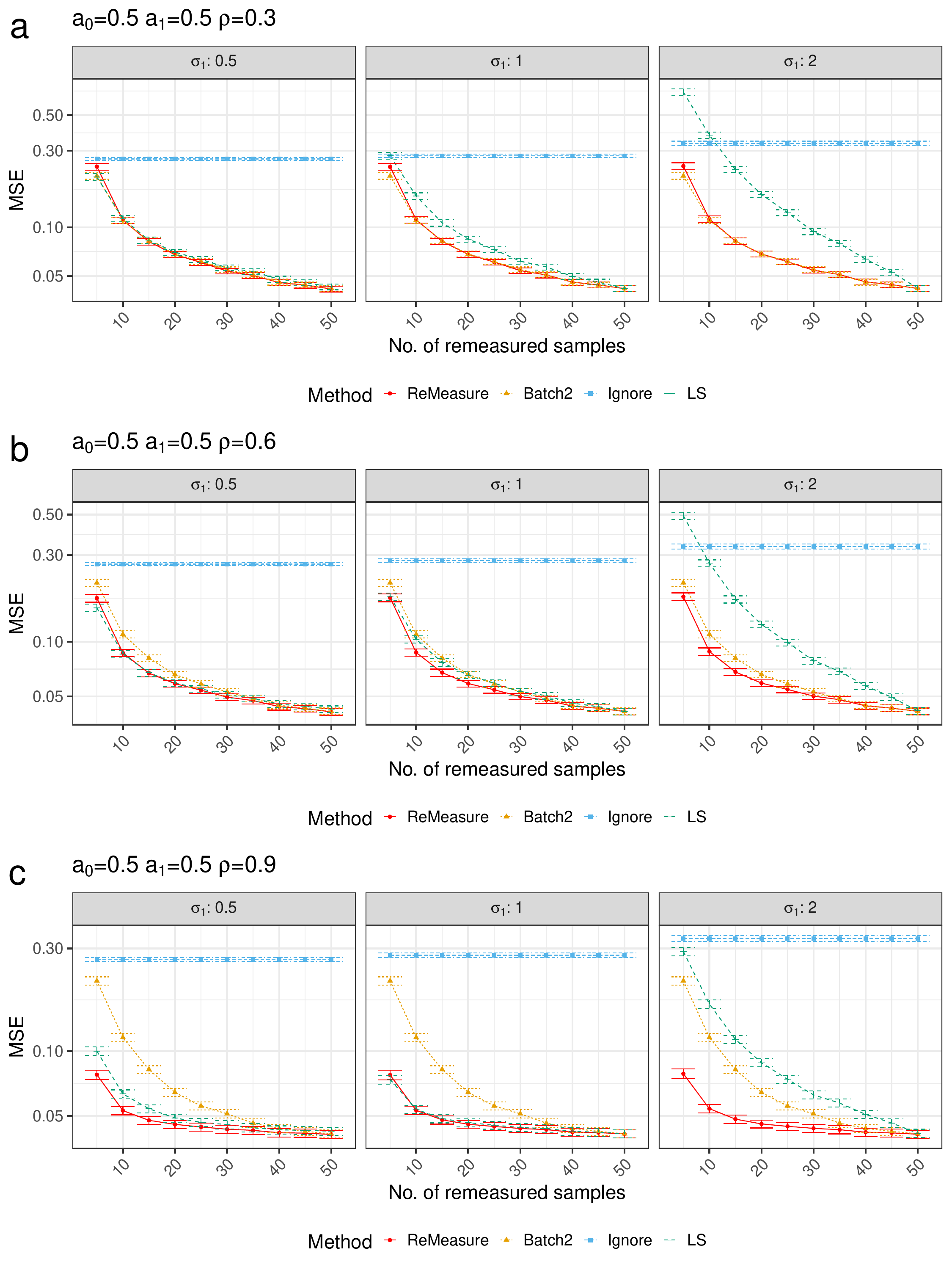}
    \caption{ {\bf The mean square error~(MSE) of $a_0$ estimate for different procedures when both sample sizes $n_1=n_2=50$.} We vary the degrees of between-batch correlation ($\rho$ values of 0.3, 0.6, and 0.9 for panels a, b, c, respectively) and degree of noise levels ($\sigma_1$, left to right). Both the biological effect parameter $a_0$  and batch location parameter $a_1$ are set to 0.5. For clarity, the y-axis is presented in $\log_{10}$ scale. Results are based on $1000$ replications. Data are presented as mean values +/- SEM.
    }
    \label{fig:MSE_S1}
    \addcontentsline{mtf}{figure}{\numberline{\thefigure}{{\bf The mean square error~(MSE) of $a_0$ estimate for different procedures when both sample sizes $n_1=n_2=50$.} We vary the degrees of between-batch correlation ($\rho$ values of 0.3, 0.6, and 0.9 for panels a, b, c, respectively) and degree of noise levels ($\sigma_1$, left to right). Both the biological effect parameter $a_0$  and batch location parameter $a_1$ are set to 0.5. For clarity, the y-axis is presented in $\log_{10}$ scale. Results are based on $1000$ replications. Data are presented as mean values +/- SEM.}} 
\end{figure}

We study the effects of location and scale differences between the two batches,  the between-batch correlations, and the number of remeasured samples. We generate the data according to the model in~\eqref{eq1}. Specifically, we set $n_1 = n_2 = 50$, and consider a univariate covariate $z_i$ randomly drawn from the standard normal distribution. We let $\vb =-0.5$ and set $\sigma_{2}^2 = 1$ so that $a_{0}$ can be interpreted as the Cohen's d~\citep{cohen2013statistical}, an effect size measure for a two-sample t-test. Let $\rho$ be the between-batch correlation for these remeasured control samples. We investigate the batch scale parameter $\sigma_{1}^2\in \{0.5^2, 1^2, 2^2\}$, the between-batch correlation $\rho \in \{0.3, 0.6, 0.9\}$, and the remeasured sample size $n_1'\in \{5, 10, 15, 20, 25, 30, 35, 40, 45, 50\}$. We set the true biological effect $a_{0} \in \{0, 0.5, 0.8\}$, representing no effect, moderate effect, and strong effect, according to Cohen's criterion. We found empirically~(Supplementary Figure~\ref{fig:Power_a1_s1}a) that the behavior of the proposed estimator of $a_{0}$ did not depend on the value of  batch location effect $a_{1}$, so we thus set $a_{1}=0.5$ throughout the simulations. 

In Figure~\ref{fig:MSE_S1} and Supplementary Table~\ref{tab:MSE_S1}, we report the mean square error (MSE) of different procedures for estimating the biological effect $a_0$ when $a_{0}=0.5$. The MSE of $a_0$ estimate for other values shows the same pattern (data not shown).
 \begin{figure}[ht!]
    \centering
    \includegraphics[width = 0.9\textwidth]{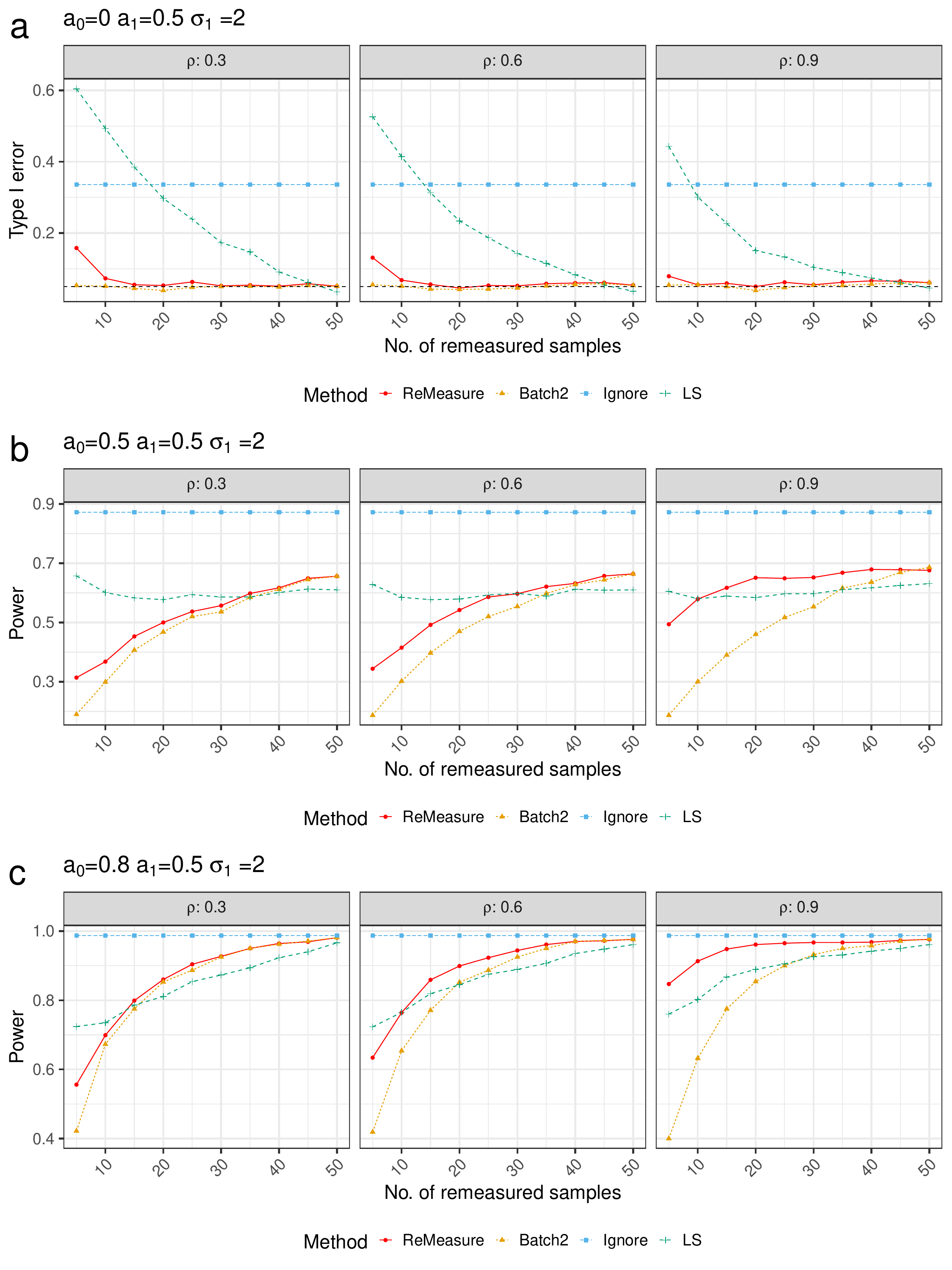}
    \caption{ {\bf Evaluation of empirical type I error and power for different procedures in testing the biological effect $a_0 = 0$ with $n_1=n_2 =50$. }  The true biological effects ($a_0$) we explored include 0 (type I error, panel a), 0.5 (power, panel b), and 0.8 (power, panel c). For each panel, from  left to right, we increase the between-batch correlation ($\rho$) from 0.3 to 0.9. The batch location parameter $a_1$ is set to 0.5 and the batch scale parameter $\sigma_1$ is set to 2. The dashed line indicates the 5\% nominal type I error rate used. }
    \label{fig:Power_S1}
    \addcontentsline{mtf}{figure}{\numberline{\thefigure}{{\bf Evaluation of empirical type I error and power for different procedures in testing the biological effect $a_0 = 0$ with $n_1=n_2 =50$. }  The true biological effects ($a_0$) we explored include 0 (type I error, panel a), 0.5 (power, panel b), and 0.8 (power, panel c). For each panel, from  left to right, we increase the between-batch correlation ($\rho$) from 0.3 to 0.9. The batch location parameter $a_1$ is set to 0.5 and the batch scale parameter $\sigma_1$ is set to 2. The dashed line indicates the 5\% nominal type I error rate used. }}
\end{figure}  
The method that ignores the batch effect and the remeasured samples (``Ignore") performs the worst in almost all settings. In contrast, the MSE for the other methods decreases with the number of remeasured samples but increases with $\sigma_{1}^2$. When the  between-batch correlation $\rho$ is small, the MSE of the method based on the second batch (``Batch2") is similar to that of the proposed method (``ReMeasure"), suggesting that the control samples in the first batch provide limited information when $\rho$ is small. In this case, using the first batch of samples may only marginally improve the estimation efficiency.    As $\rho$ becomes larger, ``Batch2" method begins to be less efficient. When the between-batch correlation is very high ($\rho = 0.9$), the control samples in the first batch help improve the estimation accuracy tremendously, and ``ReMeasure" achieves a considerably smaller MSE even when the number of remeasured samples is small. The location-scale matching method (``LS"), on the other hand, has a much higher MSE than ``ReMeasure" especially when the batch scale parameter for the first batch is large ($\sigma_1 = 2$). As the number of remeasured samples increases, the discrepancy decreases, indicating that a large number of remeasured samples may be needed for ``LS" to work properly. 

Next, we study the type I error rate and the statistical power (Figure~\ref{fig:Power_S1} ). As expected, ``Ignore" has the largest type I error inflation while ``Batch2"  controls the type I error across all settings. ``LS" has severely inflated type I error when the number of remeasured samples is small, reflecting the large MSE observed. In contrast, the proposed method ``ReMeasure" has much better type I error control than ``LS" and it generally controls the type I error to the target level when $n_1' \ge 10$. However, when the number of remeasured samples is very small ($n_1'= 5$), ``ReMeasure" has some  type I error inflation.  A larger between-batch correlation ($\rho$) reduces its inflation. The inflation is due to the use of the plug-in estimates of the variance components ($\sigma_1^2, \sigma_2^2, \rho$) in deriving the asymptotic distribution. When the number of the remeasured sample is small, the estimation of $\rho$ is subject to  large variability, and the asymptotic null distribution could deviate from the true null distribution. Indeed, if we plug in the true $\rho$ in the test statistic instead of the estimated version (``Oracle" procedure), the type I error under $n_1'= 5$ is brought down close to the target level (Supplementary Figure~\ref{fig:Theory_Oracle}a ). In terms of  statistical power, ``ReMeasure" is similar to or slightly better than ``Ignore" when $\rho$ is small 
but is substantially more powerful when $\rho$ is large. The high power of ``LS" and ``Ignore" is not very meaningful since they have severe type I error inflation.  We also compared the performance under different $\sigma_1$ values, the patterns were almost identical (Supplementary Figure~\ref{fig:Power_a1_s1}b).

To improve the type I error control under a small number of remeasured samples ($n_1'< 10$), we propose to use the bootstrap method to derive a more accurate null distribution. Supplementary Figure~\ref{fig:Power_S1_Boot} shows that the bootstrap method could control the type I error at small $n_1'$s across settings. However, the better type I error control is at the expense of some power and it is slightly less powerful than the asymptotic approach. When $\rho$ is small, it may not have any advantage over the ``Batch2" method. Therefore, the bootstrap method is only recommended for small $n_1'$s when $\rho$ is not small. 

To demonstrate the robustness of the proposed method, we performed additional simulations under large sample sizes, different batch location parameters, and different error distributions. { We also compared to two additional approaches: the naive approach, which fits a linear model based on all the samples adjusting the batch variable and ignoring repeated measurements, and the ``LSind" approach, which is the ``LS" method that uses the entire control samples to estimate the location and scale parameters.}  The results are summarized in Supplementary Section 4 (``Additional simulations").

\subsection{Theoretical Power Analysis} \label{sec:Theoretical_power}
In practice, one frequent question asked by an investigator is how many control samples need to be remeasured to achieve sufficient statistical power. Although the simulation-based approach can be used for power calculation, it is computationally intensive and is not amenable to large sample sizes. It also does not allow the exploration of different parameter settings flexibly. Therefore, an analytical power calculation tool is  needed to aid in the study design. To achieve this end, we propose an approximate power calculator based on the asymptotic distribution. Specifically, the type I error and power can be calculated theoretically through the asymptotic normality of $\ha_0$: $(\ha_0 - a_{0})/\textrm{sd}(\hat{a}_0) \Rightarrow \mathcal{N}(0,1)$. The power 
$\Pr[ |\ha_0/\textrm{sd}(\hat{a}_0)| > z_{1 - \frac{\alpha}{2}}]$ for the significant level $\alpha$ can be calculated as
\begin{equation}\label{eqn:theory}
    \Pr\left( \left| \frac{\ha_0 - a_{0} }{\textrm{sd}(\hat{a}_0)} + \frac{a_{0}}{\textrm{sd}(\hat{a}_0)}\right| > z_{1 - \frac{\alpha}{2}}\right) \approx \Pr\left(\left|Z + \frac{a_{0}}{\textrm{sd}(\hat{a}_0)}\right| > z_{1 - \frac{\alpha}{2}}\right),
\end{equation}
where $z_{1-\frac{\alpha}{2}}$ is the $(1-\alpha/2)$-quantile of the standard normal distribution and $Z\sim \mathcal{N}(0,1)$. 
In the theoretical power calculation, the oracle estimator for $\textrm{sd}(\hat{a}_0)$ is used, where we assume that $\rho, \sigma_{1}$ and  $\sigma_{2}$ are all known.

Supplementary Figure~\ref{fig:Theory_Oracle}b provides a comparison between 
the theoretical power (``Theory") and the empirical power based on the asymptotic method (``ReMeasure"). The theoretical power does not deviate much from ``ReMeasure" at different effect sizes. The approximation is more accurate when the number of remeasured samples is larger, and the between-batch correlation is higher. Thus, the theoretical power provides a reasonable approximation to the actual power when the proposed procedure is applied. 
\begin{figure}[ht]
    \centering
    \includegraphics[width = 0.9\textwidth]{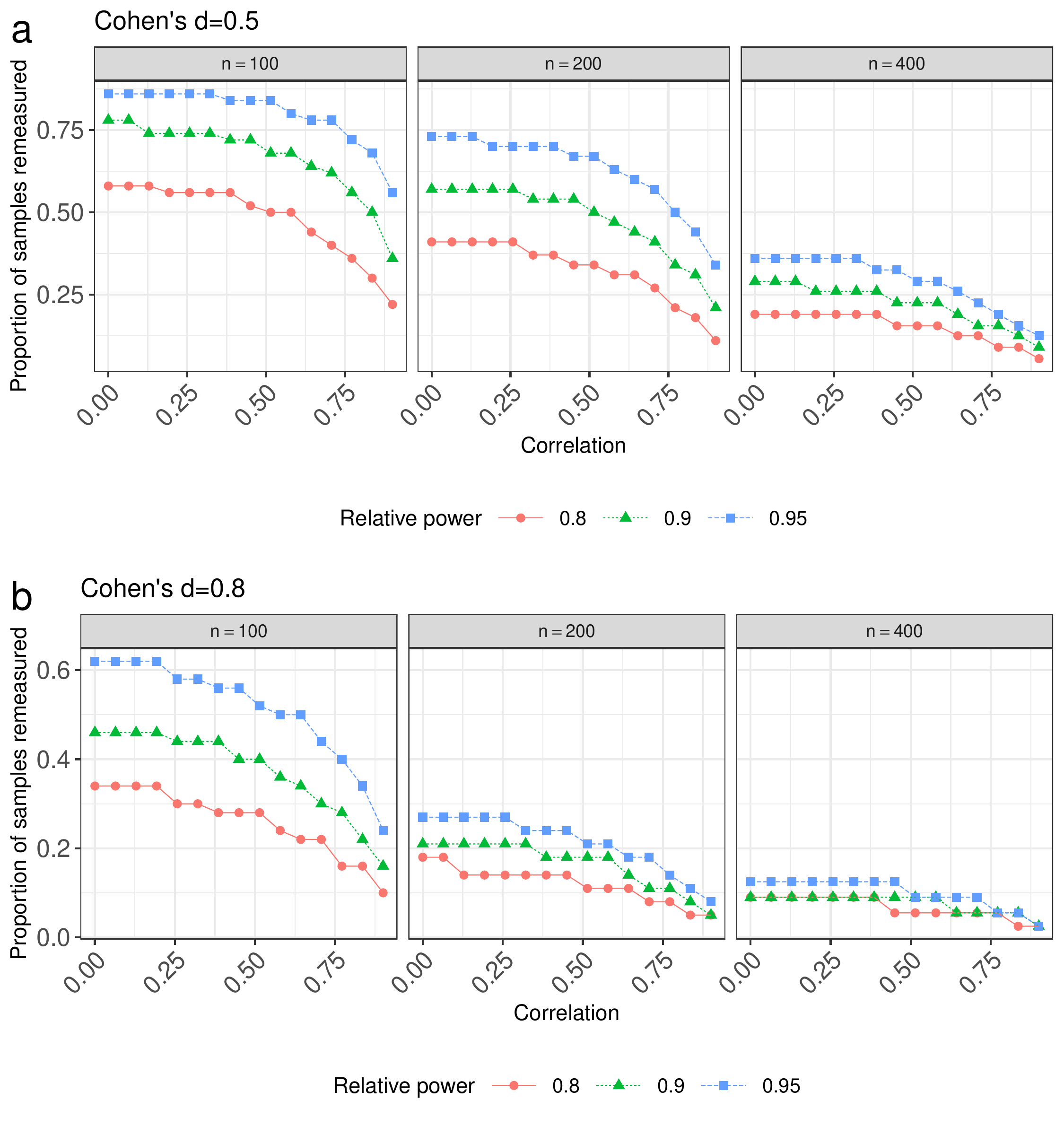}
    \caption{ {\bf Proportion of control samples that need to be remeasured to achieve $80\%, 90\%, 95\%$ relative power vs. between-batch correlation $\rho$ when $n_1 = n_2 = 50, 100, 200$.} We fix $a_1 = 0.5$ and consider settings where the effect size (Cohen's d) takes values $0.5$ (panel a) and $0.8$ (panel b), representing moderate and strong effects, respectively, according to Cohen's criterion. Results are derived from $500$ replications. 
    }
    \label{fig:Power_sample80Cor_S1}
    \addcontentsline{mtf}{figure}{\numberline{\thefigure}{{\bf Proportion of control samples that need to be remeasured to achieve $80\%, 90\%, 95\%$ relative power vs. between-batch correlation $\rho$ when $n_1 = n_2 = 50, 100, 200$.} We fix $a_1 = 0.5$ and consider settings where the effect size (Cohen's d) takes values $0.5$ (panel a) and $0.8$ (panel b), representing moderate and strong effects, respectively, according to Cohen's criterion. Results are derived from $500$ replications. }}
\end{figure} 

With the theoretical power calculator,  we can conduct power analysis  under different parameter settings. Compared to the usual parameters used in power calculation for a two-sample t-test, such as the sample size of the control group $n_1$ and the case group $n_2$, the effect size $a_0$ (Cohen's d, mean difference standardized by the within-group standard deviation), significance level, and  the desired power, power analysis for the proposed procedure depends on two additional parameters: the number of remeasured control samples $n_1'$ and the between-batch correlation $\rho$. On the other hand, the batch location and scale parameters have little effect on power. Besides traditional power analyses such as power vs. sample size and power vs. effect size, in our context, investigators are frequently interested in the following two types of power analysis:
\begin{itemize}
\item Given fixed sample sizes for the control and case group, how much power do we have at different numbers of remeasured samples?

\item Given fixed sample sizes for the control and case group, how many control samples do we need to remeasure to recover, for example, 80\% of the optimal power? The optimal power is defined as the power we can achieve by remeasuring all the control samples.
\end{itemize}

 These questions can be easily answered by the theoretical power formula. To aid in study design, we provide an R Shiny app (\url{https://hanxuan.shinyapps.io/PowerCalculation}), which takes the user-supplied parameter values (sample size, effect size, between-batch correction, significance level) as the input and  outputs the  power at different numbers of remeasured samples. We provide both the absolute and relative power, where the absolute power is the statistical power in the traditional sense, i.e., the probability of rejecting the null hypothesis when the null hypothesis is false, and the relative power is the ratio of the absolute power to the optimal power defined above.  Supplementary Figure~\ref{fig:absolute&ratio}  shows an example of power calculation for a confounded case-control study with sample remeasurement. In this example,  both the case and control sample sizes are pre-fixed at 50, the expected between-batch correlation is 0.6,
 the effect size aimed to detect (Cohen's d) is 0.6,  and the significance level used is 0.05.  The Shiny app outputs a power curve at different numbers of remeasured samples, based on which we can see that $35$ control samples need to be remeasured to achieve 80\% absolute power (Supplementary Figure~\ref{fig:absolute&ratio}a) and $19$ control samples need to be remeasured to achieve 80\% of the optimal power (Supplementary Figure~\ref{fig:absolute&ratio}b). 
 
Finally, we perform additional power analysis to gain more insights into the proposed procedure.  Figure~\ref{fig:Power_sample80Cor_S1} shows the proportion of control samples that need to be remeasured to achieve 80\%, 90\%, 95\% relative power at different sample sizes, effect sizes, and between-batch correlations. We can see that the larger the between-batch correlation, the smaller the number of samples that need to be remeasured to achieve desired relative power. The proportion of samples that need to be remeasured drops rapidly when the correlation is greater than 0.6. 

\subsection{Real Data Application} \label{sec:real_data}
We next use a real dataset to illustrate the proposed method.  The dataset came from two transcriptomics studies of ovarian cancer using different measurement platforms~\citep{wang2016expression, konecny2014prognostic,fridley2018transcriptomic}. In the first study, the gene expression was profiled using Agilent micro-arrays~\citep{konecny2014prognostic, wang2016expression}. In the second study, the gene expression was profiled using RNA-Seq~\citep{fridley2018transcriptomic}. It is well known that different measurement platform creates strong batch effects for omics study~\citep{leek2010tackling}. A subset of the samples were profiled in both studies, which provides us the opportunity to evaluate the proposed method. { In this analysis,  we focused on high-grade serous ovarian cancer, which is the most common type of ovarian cancer with well defined cancer subtypes ~\citep{chen2018consensus, konecny2014prognostic} (Agilent dataset $n=306$, RNA-Seq dataset $n=97$)}. There are 47 samples measured in both datasets. After intersecting the genes from the two 
platforms, we finally included 11,861 genes in the analysis. Based on these remeasured samples, we calculated the correlation of the gene expression between the two platforms. Supplementary Figure~\ref{fig:Power_C2C4C5_RNAseq}a shows that the distribution of the correlation coefficients has a wide range $(-0.47, 0.87)$ with a median correlation of 0.48. About 24\% genes have a correlation larger than 0.6. The overall correlation is considered to be medium.  To demonstrate the proposed method, we analyzed the cancer subtype variable (four subtypes: C1-MES, C2-IMM, C4-DIF, and C5-PRO) to identify subtype-specific gene signatures by comparing the expression profile of a specific subtype to that of the other subtypes. The Agilent dataset consists of 76, 77, 71, and 82 samples for C1-MES, C2-IMM, C4-DIF, and C5-PRO subtypes, respectively, while the RNA-Seq dataset consists of 25, 20, 28, and 24 samples for C1-MES, C2-IMM, C4-DIF, and C5-PRO subtypes, respectively.   We artificially created two sample groups with complete confounding by letting one subtype be measured on one platform and the rest three on the other platform, mimicking a completely confounded case-control study.

\begin{figure}
    \centering
    \includegraphics[width=\textwidth]{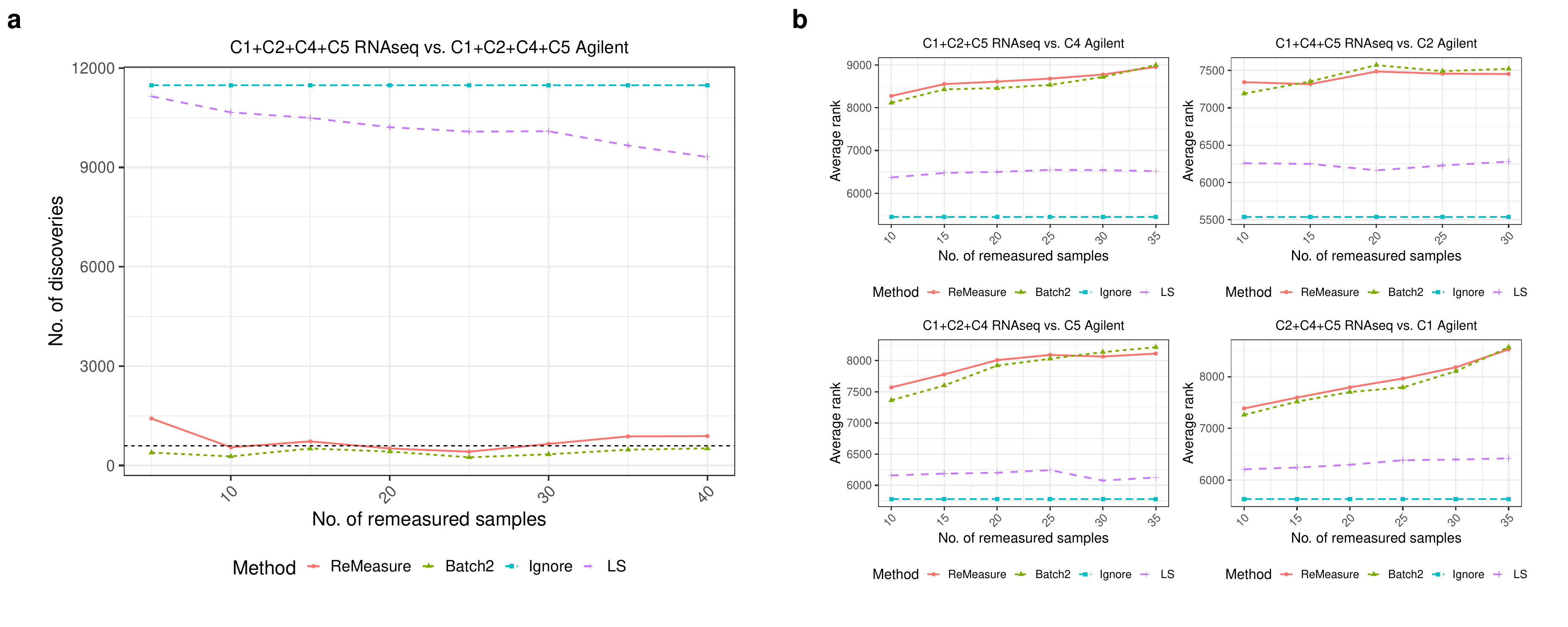}
    \caption{{\bf Comparison of ``ReMeasure", ``Batch2", ``Ignore" and ``LS" on the real dataset.} (a) The number of discoveries vs. the number of remeasured samples by comparing the same subtypes between the two platforms. Specifically, ``C1+C2+C4+C5 RNAseq vs. C1+C2+C4+C5 Agilent" denotes comparing subsets of samples, each consisting of an identical number of samples from each subtype (C1-MES, C2-IMM, C4-DIF, C5-PRO), between the RNAseq and Agilent platforms. A two-sided z-test with a 5\% significance cut-off is applied for all methods. 
    (b) The average rank vs. the number of remeasured samples for those subtype signature genes by comparing different subtypes on the two platforms. Likewise, ``C1+C2+C5 RNAseq vs. C4 Agilent"  refers to comparing combined ``C1-MES," ``C2-IMM," and ``C5-PRO" subtypes from the RNAseq platform to the ``C4-DIF" subtype from the Agilent platform. The same explanation applies to other titles. }
    \label{fig: Discover_rank}
\end{figure}

We first compare the performance of ``ReMeasure", ``Batch2", ``Ignore" and ``LS" after fitting gene-wise models. We start with evaluating the type I error control of the proposed method. This is achieved by comparing the same subtypes from the Agilent and the RNA-Seq platform. To ensure sufficient statistical power, we pooled samples from all four subtypes and made the subtype composition similar between the Agilent and RNA-Seq dataset. Specifically, we compare $276$ Agilent samples consisting of 69 samples in each subtype to $68$ RNA-Seq samples consisting of $17$ samples in each subtype, using $40$ remeasured Agilent samples to correct for batch effects.  Since both batches have similar subtype composition and the patient characteristics are also similar between the two batches (they are from the same Midwest population), we expect to see very few substantial differences. 
Indeed, based on Figure~\ref{fig: Discover_rank}, we observe that ``Batch2" detects about 5\% ``significant'' genes across different numbers of remeasured samples as expected at the 5\% significance cutoff.  For ``ReMeasure", it  detects close to 5\%  ``significant'' genes  when the number of remeasures samples are larger than or equal to 10, consistent with the simulation results.   In contrast, ``Ignore" and ``LS" have made substantially more ``false" discoveries, indicating that they have poor type I error control.
\addcontentsline{mtf}{figure}{\numberline{\thefigure}{{\bf Comparison of ``ReMeasure", ``Batch2", ``Ignore" and ``LS" on the real dataset.} (a) The number of discoveries vs. the number of remeasured samples by comparing the same subtypes between the two platforms. Specifically, ``C1+C2+C4+C5 RNAseq vs. C1+C2+C4+C5 Agilent" denotes comparing subsets of samples, each consisting of an identical number of samples from each subtype (C1-MES, C2-IMM, C4-DIF, C5-PRO), between the RNAseq and Agilent platforms. A two-sided z-test with a 5\% significance cut-off is applied for all methods. 
    (b) The average rank vs. the number of remeasured samples for those subtype signature genes by comparing different subtypes on the two platforms. Likewise, ``C1+C2+C5 RNAseq vs. C4 Agilent"  refers to comparing combined ``C1-MES," ``C2-IMM," and ``C5-PRO" subtypes from the RNAseq platform to the ``C4-DIF" subtype from the Agilent platform. The same explanation applies to other titles. }}   

Next, we conduct a power study by comparing one subtype from the Agilent platform to the other three subtypes from the RNA-Seq platform, treating the RNA-Seq samples as controls and the Agilent samples as cases.  RNA-Seq samples remeasured on the Agilent platform are used to correct batch effects.  To objectively evaluate power, we need to know the ground truth. However, the ground truth is unknown in this case, so instead we create a list of genes that are more likely to be subtype signatures by comparing one subtype vs. others in the same Agilent dataset. Based on two-sample t-tests and 5\% FDR~(Benjamini-Hochberg procedure), we identified $3793, 4212, 6168, 4439$ signature genes for the four subtypes, respectively.  
In the following, we conduct four types of comparisons: 1) C1+C2+C5 RNA-Seq vs. C4 Agilent, 2)  C1+C4+C5 RNA-Seq vs. C2 Agilent,  3)  C1+C2+C4 RNA-Seq vs. C5 Agilent and 4)  C2+C4+C5 RNA-Seq vs. C1 Agilent.  We evaluate the ability of the proposed method to retrieve those signature genes, in comparison to ``Batch2", ``LS" and ``Ignore".  If a method works,  we expect that the signature genes will rank high (lower p-values) in the respective results. 
The average ranks of ``ReMeasure" and ``Batch2" are much higher than ``Ignore" and ``LS" 
(Figure~\ref{fig: Discover_rank}).``ReMeasure" achieves a slightly higher rank than ``Batch2", especially when the number of remeasured samples is at the lower end. However, ``ReMeasure" recovers substantially more  genes than ``Batch2" for the four subtypes at 5\% FDR  (Supplementary Figure~\ref{fig:Power_C2C4C5_RNAseq}b),
indicating that ``ReMeasure" is more powerful than ``Batch2" while the false positive control is similar to ``Batch2".

Finally, we compare  the number of discoveries for ``Batch2" and ``ReMeasure" on the genes with the lowest between-batch correlation (bottom quartile) and the highest between-batch correlation (top quartile). 
Supplementary Figures~\ref{fig:Power_C2C4C5_RNAseq}c 
and~\ref{fig:Power_C2C4C5_RNAseq}d reveal that our approach is more similar to ``Batch2" under weak correlation and more powerful than  ``Batch2" under a strong correlation, consistent with our simulation study.
\begin{figure}[ht!]
    \centering
    \includegraphics[width=\textwidth]{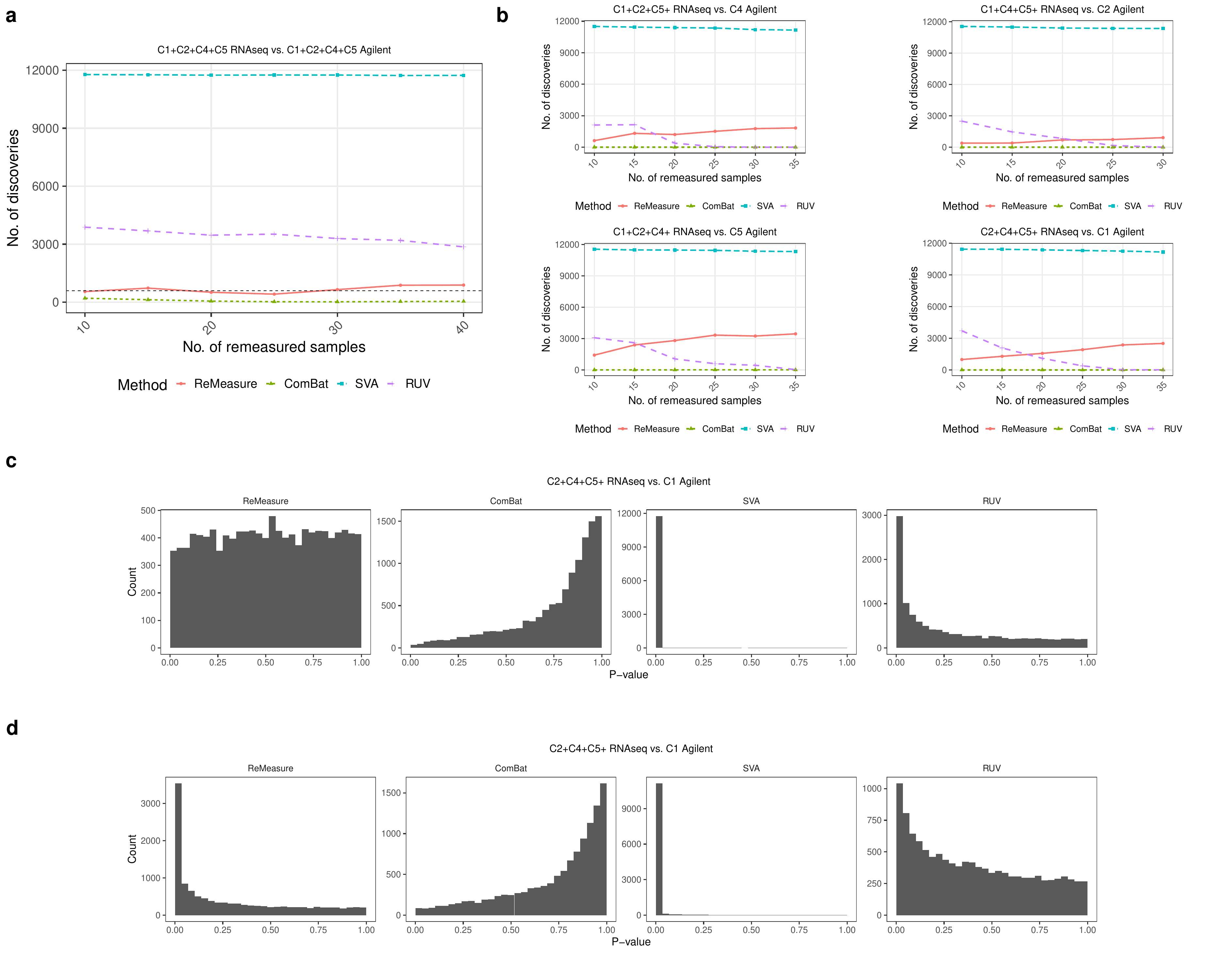}
    \caption{{\bf Comparison to ``ComBat", ``SVA" and ``RUV" on the real dataset.}  (a) and (b) show the number of discoveries vs. the number of remeasured samples by comparing subtypes on the two platforms. The p-values of ``ComBat", ``SUV" are calculated based on the F-test, while those of ``RUV" and ``ReMeasure" are obtained from two-sides t-test and z-test, respectively. 
    (a) Comparing the same subtypes when the nominal type I error level is 0.05. ``C1+C2+C4+C5 RNAseq vs. C1+C2+C4+C5 Agilent" denotes comparing subsets of samples, each consisting of an identical number of samples from each subtype (C1-MES, C2-IMM, C4-DIF, C5-PRO), between the RNAseq and Agilent platforms. A 5\% significance cut-off is applied for all methods. (b) Comparing different subtypes with 5\% FDR. ``C1+C2+C5 RNAseq vs. C4 Agilent"  refers to comparing combined ``C1-MES," ``C2-IMM," and ``C5-PRO" subtypes from the RNAseq platform to the ``C4-DIF" subtype from the Agilent platform. The same explanation applies to other titles. 
    (c) and (d) show the unadjusted raw p-value distribution. (c) Comparing the same subtypes  with all $40$ remeasured samples included. (d) Comparing different subtypes (C2+C4+C5 RNA-Seq vs. C1 Agilent) with all $35$ remeasured samples included.  }
    \label{fig:SvaComBatRUV}
\end{figure}

\addcontentsline{mtf}{figure}{\numberline{\thefigure}{{\bf Comparison to ``ComBat", ``SVA" and ``RUV" on the real dataset.}  (a) and (b) show the number of discoveries vs. the number of remeasured samples by comparing subtypes on the two platforms. The p-values of ``ComBat", ``SUV" are calculated based on the F-test, while those of ``RUV" and ``ReMeasure" are obtained from two-sides t-test and z-test, respectively. (a) Comparing the same subtypes when the nominal type I error level is 0.05. ``C1+C2+C4+C5 RNAseq vs. C1+C2+C4+C5 Agilent" denotes comparing subsets of samples, each consisting of an identical number of samples from each subtype (C1-MES, C2-IMM, C4-DIF, C5-PRO), between the RNAseq and Agilent platforms. A 5\% significance cut-off is applied for all methods. (b) Comparing different subtypes with 5\% FDR. ``C1+C2+C5 RNAseq vs. C4 Agilent"  refers to comparing combined ``C1-MES," ``C2-IMM," and ``C5-PRO" subtypes from the RNAseq platform to the ``C4-DIF" subtype from the Agilent platform. The same explanation applies to other titles. 
    (c) and (d) show the unadjusted raw p-value distribution. (c) Comparing the same subtypes  with all $40$ remeasured samples included. (d) Comparing different subtypes (C2+C4+C5 RNA-Seq vs. C1 Agilent) with all $35$ remeasured samples included.} }  


We further compare our ``ReMeasure'' method to ``ComBat"~\citep{johnson2007adjusting},  ``SVA"~\citep{leek2007capturing, leek08}, and ``RUV"~\citep{gagnon2012using,gagnon2013removing,jacob2016correcting}, the three most popular batch effect correction methods, on the real data set.
``ComBat" directly removes the known batch effects by performing an empirical Bayesian adjustment, while  ``SVA" identifies and estimates the surrogate variables for unwanted variations, including batch effects and other unmeasured biological variations, with no requirement of knowing the batch a sample belongs to. {  ``RUV" assumes a factor model that utilizes negative control genes~(i.e., genes unrelated to the factor of interest) to estimate the latent factors for unwanted variations. } Although in our case the batch information is known, we still run ``SVA" and ``RUV" to see whether they can capture the known batch effects. We used the \texttt{ComBat} and \texttt{sva}  functions in the R Bioconductor \texttt{sva} package, { and \texttt{naiveReplicateRUV} function in the R Bioconductor \texttt{RUVnormalize} package } to run the three procedures. The remeasured samples in the second batch were included in the analysis, but their corresponding samples in the first batch were excluded to satisfy the independence assumption of both methods. For ``SVA", we used the permutation method described in~\cite{buja1992remarks} to estimate the optimal number of surrogate variables. The resulting surrogate variables were then included in the regression model as  covariates. The p-values were calculated based on the F-test, comparing the model with and without the group variable. For ``ComBat", we  fit the gene-wise linear regression model based on batch-corrected data. { For ``RUV", 364 housekeeping genes were used as negative controls, and the remeasured samples were used as the replicates, following the original paper~\citep{jacob2016correcting}.  }

 Under the null, where we compare the gene expression of the same subtypes (C1+C2+C4+C5) between the two measurement platforms (RNA-Seq vs. Agilent),   ``SVA" finds a substantially higher number of significant genes than what would be predicted under the null, even with a large number of estimated surrogate variables  ($>$24 surrogate variables for most cases, Figure~\ref{fig:SvaComBatRUV}a), indicating that the estimated surrogate variables are still not adequate to capture the full batch effects. Since ``SVA" could not control the type I error properly,  its high power under the alternative hypothesis  is thus not meaningful (Figure~\ref{fig:SvaComBatRUV}b).  On the other hand,   ``ComBat" is very conservative and finds very few significant genes under the null (Figure~\ref{fig:SvaComBatRUV}a). Its type I error control is at the expense of power.  When we compare one subtype vs. others (Figure~\ref{fig:SvaComBatRUV}b), the power of ``ComBat" is extremely low, indicating that most of the true biological signals may be removed in batch correction due to high confounding of biological and batch effects. { ``RUV" also has substantially
increased type I error, but is less serious than ``SVA". The number of detected significant genes decreases with the number of remeasured samples.  }

 It is also interesting to compare the p-value distributions of the four methods. Under the null (C1+C2+C4+C5 RNA-Seq vs. C1+C2+C4+C5 Agilent), the p-value distribution of ``ReMeasure" is close to the uniform distribution, while the p-value distributions of ``ComBat", ``SVA", and ``RUV" deviate substantially from the uniform distribution (Figure~\ref{fig:SvaComBatRUV}c).  When comparing C2+C4+C5 RNA-Seq to C1 Agilent (Figure~\ref{fig:SvaComBatRUV}d), the p-value distribution of ``ReMeasure" has the expected form for a multiple testing experiment with signals, with a spike of small p-values and a long tail of larger p-values close to the uniform distribution. In contrast,  the p-value distribution of ``ComBat" has a spike on the right side of the histogram due to over-adjustment, and the p-values of ``SVA" concentrate on the left side of the histogram due to under-adjustment. The p-value distribution of ``RUV"  behaves well in this case.
 
 We thus conclude that the existing batch adjustment methods do not work well in the severely confounded scenario, and our method can effectively leverage the remeasured samples to correct batch effects.

\section{Discussion}

 Due to the complex technical processes involved in biological measurement, even slight variation in sample preparation and processing can cause batch effects~\citep{leek2010tackling}. In many cases, batch effects are not known until the data are analyzed. Batch effects are most disastrous when it is highly confounded with the variable of interest, for example, when the case and control samples are measured separately.  In such scenarios, it is extremely challenging to separate the true biological effects from batch effects. Although such confounded studies could be due to a bad study design or less awareness of batch effects, they could also be due to logistics issues.  For example, a clinical investigator has collected patient samples and wants to compare them to existing controls.  But due to sample availability or financial constraint,  the investigator may not be able to remeasure all the control samples together with the case samples. It is thus of tremendous help to the investigator if she only needs to remeasure a small subset of control samples while retaining most power. 
 
 Traditional batch effect correction methods such as ``ComBat" \citep{johnson2007adjusting}, ``SVA"~\citep{leek2007capturing, leek08}, and ``RUV"~\citep{gagnon2012using,gagnon2013removing,jacob2016correcting} were mainly developed for independent samples and they have limited ability to correct batch effects in highly confounded scenario. They either removed the batch and biological effects altogether (reduced power) or retained the batch effects to a large extent (increased type I error).  
 
 Our method has several limitations. In some cases, the control samples may not be available for remeasurement, making our method not applicable. Even if they are available for remeasurement, there can still be subtle batch effects associated with difference in collection,  storage, and freeze-thaw cycle \citep{rundle2005design}.  Though reprocessing  the samples can reduce batch effects, batch effects associated with the upstream technical variation can still persist. Our method cannot correct these residual batch effects. Furthermore, although we show our method is robust to some deviation of the Gaussian distribution, it can still perform poorly when the data are highly skewed or zero inflated. As the genomics studies move into the era of single-cell genomics, the genomics data has become even more complex with severe zero inflation \citep{stegle2015computational}. Simple data transformation may not be sufficient to make the data Gaussian-like. To extend the capability of our method to analyze such complex genomics datasets, new methodological development is needed.  One potential direction is to extend our method to the generalized linear model setting, where the measurement can be modeled by more general distributions such as zero-inflated negative binomial model for zero-inflated count data \citep{chen2018omnibus, risso2018general}.  
 
Our procedure is based  on the maximum likelihood estimation framework, and we  proved its consistency and asymptotic normality. However, when the number of remeasured samples is small  ($n{<}10$), the procedure could have inflated type I error.   This is a disadvantage of the proposed method since the number of samples needed to be remeasured may be small when the between-batch correlation is high. To improve the small-sample performance, we proposed a bootstrap method based on residual resampling and showed that it had a well-controlled type I error. { However, when the inter-batch correlation is not high (${<}0.8$),  the bootstrap method could be less powerful than the ``Batch2" method.  In this case, ``Batch2" is recommended. } As the type I error inflation of the asymptotic procedure is mainly driven by the inaccurate estimation of the between-batch correlation when we analyze a large number of features as in omics-wide testing,  it is possible to improve the estimation efficiency by pooling information from all features  using empirical Bayes method~\citep{johnson2007adjusting}. We leave this as a future research direction. 

\section{Methods}\label{sec:Method}

\subsection{Parameter Estimation}\label{subsec:Estimate_para}
Under the Gaussian assumption on the errors, the log joint likelihood of the data is given by
\begin{equation}
    { \footnotesize 
    \begin{aligned}
        \label{eqn:objective_S1}
   & L_{\mN}(\btheta)  = 
    -  n_1' \log(\sigma_1^2) - n_1' \log(\sigma_2^2) - n_1' \log(1 - \rho^2) \\
    - &\frac{1}{(1 - \rho^2)} \sum_{i \in S_1} \Bigg[ \left(\frac{y_i - \mu_{1i}}{\sigma_1}\right)^2 - 2\rho\left(\frac{y_i - \mu_{1i}}{\sigma_1}\right)\left(\frac{y_{n+i} - \mu_{3i}}{\sigma_2}\right) + \left(\frac{y_{n+i} - \mu_{3i} }{\sigma_2}\right)^2 \Bigg] \\
   - & (n_1 - n_1')\log(\sigma_1^2) - \sum_{i \in C_1\setminus S_1} \left(\frac{y_i - \mu_{1i}}{\sigma_1}\right)^2 -  n_2\log(\sigma_2^2) -  \sum_{i \in T_2} \left(\frac{y_i - \mu_{2i}}{\sigma_2}\right)^2.
    \end{aligned}
    }
\end{equation}
The maximum likelihood estimator (MLE) of $\btheta$ can be obtained as 
\begin{equation}\label{eq:mle}
    \begin{split}
    \hat{\btheta}=(\hat{a}_0, \hat{a}_1, \hat{\vb}, \hat{\rho}, \hat{\sigma}_1, \hat{\sigma}_2) = \argmax_{\btheta \in \Theta} L_{\mN}(\btheta).
    \end{split}
\end{equation}
The solution to (\ref{eq:mle}) does not have a closed-form solution due to the correlation between the remeasured samples from the two batches. One way to find the solution is by using a generic numerical optimization algorithm such as the Newton-Raphson method or its variants, which updates the parameters via the first or second-order methods until convergence. Here we provide a more efficient algorithm (see Supplementary Section 2) that explores the specific structure of the first-order conditions associated with the objective function. We deduce the first-order conditions by setting the partial derivative of the objective function with respect to each parameter to be zero. We then update the parameters by iteratively solving these equations. The algorithm is an order-of-magnitude faster than the generic optimization algorithm ({Supplementary Figure~\ref{fig:Time_S1}}). 

\subsection{Statistical Inference}\label{subsec:Inference}
We are mostly interested in estimating and conducting inference of the biological effect $a_0$. The uncertainty assessment or variance of $\hat{a}_0$ is key to hypothesis testing and power analysis. The alternate updating algorithm in Supplementary Section 2 allows us to obtain the variance estimate straightforwardly. It can be shown that
\begin{equation}
    \hat{a}_0 = \sum_{i \in T_2} \frac{y_i - \z_i^\top \hat{\vb} }{n_2} - \sum_{i \in C_2 } \frac{y_i - \z_i^\top \hat{\vb} }{n_1'} +  \frac{\hat{\rho}\hat{\sigma}_2}{\hat{\sigma}_1} \sum_{i\in S_1} \frac{y_i - \z_i^\top \hat{\vb}}{n_1'} \label{form:a0}
\end{equation}
indicating that the MLE of $a_0$ can be expressed as the linear combination of  response variables from different batches. The first two terms in the formula estimate the biological effect without using the first batch of samples, and the third term uses those remeasured samples in the first batch to adjust the estimate. The degree of adjustment  depends on the  between-batch correlation for those remeasured samples. When the correlation is low, the estimate is similar to that without using the first batch. However, when the correlation is high, the adjustment could be substantial.  Based on the formula (\ref{form:a0}), we can calculate its variance accordingly. The details for the variance formula can be found in Supplementary Section 2. 

Based on the large-sample theory in Supplementary Section~\ref{apx:sec:proofs}, the p-value for testing $a_0=0$ can be computed as $2\Phi(-|\hat{a}_0|/\widehat{\textrm{sd}}(\hat{a}_0)),$ where $\Phi(\cdot)$ denotes the cumulative distribution function of the standard normal distribution. The remeasured sample size $n_1'$ needs to be large for our large-sample theory to work. However, in practice, the remeasured sample size may be small, in which case, the estimation of the correlation parameter $\rho$ is subject to large variability since it only depends on $n_1'$ pairs of observations. For small $n_1'$, the large sample theory does not provide an accurate approximation to the sampling distribution of $|\hat{a}_0|/\widehat{\textrm{sd}}(\hat{a}_0)$.
To overcome this issue, we propose to use the bootstrap method to improve the approximation accuracy of the finite sample distribution. For example, we can use the residual bootstrap. The set of residuals is obtained as  
\begin{align*}
    &\text{Control (batch 1):} \quad \hat{\epsilon}_i^{(1)} = y_i - \z_i^\top \hat{\b}, i = 1, \ldots, n_1 ,\\
    &\text{Case (batch 2):} \quad \hat{\epsilon}_i^{(2)} = y_i - \hat{a}_0 - \hat{a}_1  - \z_i^\top \hat{\b}, i = n_1 +1, \ldots, n_1+n_2 = n,\\
    &\text{Control (batch 2):} \quad \hat{\epsilon}_i^{(2)} = y_i - \hat{a}_1 -  \z_i^\top \hat{\b}, i = n+1, \ldots, n+n_1'. 
\end{align*}
We re-sample the residuals with replacements from each group and then generate a new bootstrap sample with the fixed $\z_i$ but new $y_i$ using the fitted parameters and re-sampled residuals. %

Given $B$ bootstrap samples, we can calculate $\hat{a}_0^{(b)}$ and $\widehat{\Var}(\hat{a}_0^{(b)})$ for $1\leq b\leq B$ based on each resample using Algorithm \ref{algorithm:S1} and Formula (\ref{eq-var}) in Supplementary Section 2. Thereby, we obtain the bootstrap statistics $Z_b:= (\hat{a}_0^{(b)} - \hat{a}_0)/\sqrt{\widehat{\Var}(\hat{a}_0^{(b)})}$ for $b = 1,\ldots,B$. Given $Z = \hat{a}_0/\sqrt{\widehat{\Var}(\hat{a}_0)}$, the bootstrapped p-value can be computed as $B^{-1}\sum_{b=1}^B \v1\{ |Z_b| > |Z| \}.$

\backmatter



\bmhead{Data Availability} 
Source data for Figures 2-6 is available with this manuscript.  They can also be found at \url{https://github.com/yehanxuan/BatchReMeasure-manuscript-sourcecode}. 

\bmhead{Code Availability} All the codes to reproduce the results in this paper are available at \url{https://github.com/yehanxuan/BatchReMeasure-manuscript-sourcecode}. The developed R package BatchReMeasure is available at \url{https://github.com/yehanxuan/BatchReMeasure}. The specific version used to produce the results in this manuscript is also available on Code Ocean~\citep{CodeOceanYe}.







\begin{appendices}

\renewcommand{\thelemma}{{S.\arabic{lemma}}}
\renewcommand{\figurename}{Supplementary Fig.} 
\renewcommand{\thefigure}{\arabic{figure}}
\setcounter{table}{0}
\renewcommand{\tablename}{Supplementary Table} 
\renewcommand{\thetable}{\arabic{table}}
\renewcommand{\thesection}{\arabic{section}}

\section{Theoretical results} \label{apx:sec:proofs}

This section will show that our estimator is consistent and asymptotically normal if the sample size grows to infinity. Our theory differs from the traditional MLE theory in two aspects: (i) we do not require the likelihood function to be correctly specified as the errors are allowed to be non-Gaussian, and (ii) the data are not identically distributed in our setting as the model structure changes across batches and case/control groups, which complicates the analysis. The following set of mild assumptions is imposed for theoretical analysis.
\begin{assumption} \label{ass: compact}
  The true parameter $\btheta_0 = (a_{00}, a_{10}, \vb_0, \rho_0, \sigma_{10}, \sigma_{20})$ belongs to the interior of some compact parameter space $\Theta$. 
\end{assumption}
\begin{assumption} \label{ass: moment}
The errors $\epsilon_{i}^{(j)}$ are independent across $i$ and $j$ with mean zero and finite variance, i.e.,
$\Expect[\epsilon_i^{(j)}] = 0$ and $\Var[\epsilon_i^{(j)}] = \sigma_j^{2} < \infty$. Assume that the covariates $\vz_i$ are i.i.d with $\Expect[\vz] = \mathbf{0}$ and $\Expect[\vz \vz^\top] = \mSigma_z$, where $\mSigma_z$ is positive definite.  
\end{assumption}
\begin{assumption} \label{ass:portion}
    Suppose each batch is a non-negligible portion of the total sample, and the remeasured sample is a non-negligible portion of the batch 1 sample. Formally, we assume that $ n_1/n \rightarrow r_1$ and $ n_1'/n_1 \rightarrow r_1'$ as $n_1, n_2 \rightarrow \infty$, where $r_1,r_1'\in (0,1)$. 
\end{assumption}

Under Conditions~\ref{ass: compact}-\ref{ass:portion}, the objective function normalized by the sample size converges in probability to a weighted sum of some non-stochastic functions as $n_1', n_2 \rightarrow \infty$ by the law of large numbers,
\begin{equation} \label{eqn:gen_model}
    \begin{aligned} 
    \bar{L}_{\mN}(\btheta) = n^{-1}L_{\mN}(\btheta)  \stackrel{p}{\rightarrow} \ell(\btheta) = \sum_{k=1}^3 w_k \ell_k (\btheta)
\end{aligned}
\end{equation}
with $w_1 = r_1r_1', w_2 = r_1(1-r_1')$ and $w_3 = (1-r_1)$. Here $\ell_{k}(\btheta)$'s are the limiting functions of the sample averages of the Gaussian log-likelihoods. The detailed forms of $\ell_{k}(\btheta)$ can be found in Section~\ref{apx:thm:consistency}. 

\begin{theorem}[Consistency] \label{thm:consistency}
Suppose Conditions~\ref{ass: compact}-\ref{ass:portion} are satisfied. The estimator $\hat{\btheta}$ that maximizes the objective function $L_{\mN}(\btheta)$ is weakly consistent, namely,  $\hat{\btheta}$ converges in probability to the underlying true parameter $\btheta_0 = (a_{00}, a_{10}, \vb_0, \rho_0, \sigma_{10}, \sigma_{20})$ of Model~\ref{eq1} as $n_1' \rightarrow \infty, n_2 \rightarrow \infty$.
\end{theorem}
We remark that the remeasured size $n_1'$ has to tend to infinity to ensure the consistency of the MLE of $\rho$.
\begin{theorem}[Asymptotic Normality]
\label{thm: asymptotic}
    Under Conditions~\ref{ass: compact}-\ref{ass:portion}, 
    the estimator $\hat{\btheta}$ is asymptotically normal, i.e.,
\begin{equation}\label{eqn:asymptotic}
    \begin{aligned} 
         \sqrt{n} (
         \hat{\btheta}- \btheta_0 )  \Longrightarrow \mathcal{N}\left(0,\Expect[\ddot{\ell}(\btheta_0)]^{-1} \Expect\left[ \sum_{k=1}^3 w_k \dot{\ell}_{k}(\btheta_0) \dot{\ell}_{k}(\btheta_0)^\top \right] \Expect[\ddot{\ell}(\btheta_0)]^{-1}\right),
    \end{aligned}
\end{equation}
as $n_1', n_2 \rightarrow \infty$, with the weights $w_k$'s given in~\eqref{eqn:gen_model}. Here $\dot{\ell}_{k}(\btheta_0)$ and $\ddot{\ell}(\btheta_0)$ denote the first derivative of $\ell_k(\btheta)$and second derivative of $\ell(\btheta)$ at $\btheta=\btheta_0$, respectively. 
\end{theorem}

Proofs of Theorem \ref{thm:consistency} and \ref{thm: asymptotic} are detailed below.


\subsection{Proof of Theorem~\ref{thm:consistency}} \label{apx:thm:consistency} 

The proof requires multiple steps. We first present several useful lemmas. Then, we show that the objective function converges uniformly in probability to some non-stochastic function that has a unique maximizer. The consistency is then established using Lemma \ref{lem:takeshi}.

\begin{lemma}[Strictly concavity]
  The log-likelihood of a mean-zero Gaussian distribution
  \begin{equation} \label{eqn: logdet}
    h_n(\mSigma^{-1}) =-\frac{1}{2}\text{tr}{ (\mS_n \mSigma^{-1})}  + \frac{1}{2}\log \det(\mSigma^{-1}) 
  \end{equation}
  is strictly concave with respect to $\mSigma^{-1}$ for some positive definite matrix $\mS_n$. Thus $h_n(\cdot)$ has a unique global maximizer.
\end{lemma}
\begin{proof}
Let $\mOmega = \mSigma^{-1}$. Note that $-\tr(\mS_n \mOmega)$ is an affine function of $\mOmega$, and the log-determinant function $\log \det(\mOmega)$ is strictly concave. Thus the linear combination of these two terms is strictly concave as a function of $\mOmega$.
\end{proof}
\begin{lemma}[Theorem 4.1.1 of~\cite{takeshi1985advanced}] \label{lem:takeshi} 
  Suppose the function $\bar{L}_{\mN}(\btheta)$ satisfies the following conditions:
  \begin{enumerate}
      \item The parameter space $\Theta$ is compact. 
      \item $\bar{L}_{\mN}(\btheta)$ is continuous in $\btheta \in \Theta$ almost everywhere. 
      \item $\bar{L}_{\mN}(\btheta)$ converges to a non-stochastic function $\ell(\btheta)$ in probability uniformly over $\btheta \in \Theta$ and $\ell(\btheta)$ attains a unique global maximum at $\btheta_0$.
  \end{enumerate}
  Then $\hat{\btheta}_{\mN} := \argmax_{\btheta \in \Theta} \bar{L}_{\mN}(\btheta)\rightarrow^p \btheta_0.$
\end{lemma}
\begin{lemma}[Uniform convergence in probability]\label{lem:uniform}
Let $g(\vx, \btheta)$ be a measurable function of $\vx$ for each $\btheta$ in a compact space $\Theta$, and a continuous function of $\btheta$ for each $\vx$. Let $\vx_{i}$ be a sequence of i.i.d random vectors such that $\Expect[\sup_{\btheta \in \Theta}|g(\vx_i, \btheta)| ]< \infty$
 and $\Expect[g(\vx_i, \btheta)]=0.$
Then 
$$
\frac{1}{n} \sum_{i=1}^n g(\vx_i, \btheta) \rightarrow^p 0 \quad \mbox{ uniformly.}
$$
\end{lemma}
\begin{proof}
 Write $g_i(\btheta) = g(\vx_i, \btheta)$ for the ease of notation. The compact parameter space $\Theta$ has a finite non-overlapping cover $\Theta_1^K, \ldots, \Theta_K^K$ such that the distance of any two points within some $\Theta_i^K$ goes to $0$ as $K \rightarrow \infty$. Let $\btheta_1, \ldots, \btheta_K$ be $K$-vectors such that $\btheta_i \in \Theta_i^K$. Then we have for any $\varepsilon > 0$, 
\begin{align*}
    & \Pr[\sup_{\btheta \in \Theta} |n^{-1}\sum_{i=1}^{n}g_{i}(\btheta)| > \varepsilon] \le  \Pr[\cup_{k=1}^{K} \{ \sup_{\btheta \in \Theta_k^K } |n^{-1}\sum_{i=1}^{n}g_{i}(\btheta)| > \varepsilon \} ] \\
    \le & \sum_{k=1}^K \Pr[ \sup_{\btheta \in \Theta_k^K } |n^{-1}\sum_{i=1}^{n}g_{i}(\btheta)| > \varepsilon ] \\
    \le & \sum_{k=1}^K \Pr[  |n^{-1}\sum_{i=1}^{n}g_{i}(\btheta_k)| > \varepsilon/2 ] + \sum_{k=1}^K \Pr[  n^{-1}\sum_{i=1}^{n} \sup_{\btheta \in \Theta_k^K } |g_{i}(\btheta) - g_{i}(\btheta_k)| > \varepsilon/2 ].
\end{align*}
Since $g_{i}(\btheta)$ is uniformly continuous in $\btheta \in \Theta$ for every $i$, we have
$$
\lim_{K \rightarrow \infty} \sup_{1\leq k\leq K} \sup_{\btheta \in \Theta_k^K } |g_i(\btheta) - g_{i}(\btheta_k) | = 0
$$
almost surely. Meanwhile, 
\begin{equation}
\label{eqn:supbound}
 \sup_{1\leq k\leq K} \sup_{\btheta \in \Theta_k^K } |g_i(\btheta) - g_{i}(\btheta_k) | \le 2 \sup_{\btheta \in \Theta}|g_i(\btheta)|.    
\end{equation}
The integrability of the right-hand side indicates that we can use the Lebesgue dominated convergence theorem to show that
$$
\lim_{K \rightarrow \infty} \Expect[\sup_{1\leq k\leq K} \sup_{\btheta \in \Theta_k^K}|g_{i}(\btheta) - g_{i}(\btheta_k)| ] = 0.
$$
That is to say, there exists a finite $K= K(\varepsilon/4)$ such that
$$
\Expect[\sup_{\btheta \in \Theta_k^{K}}|g_{i}(\btheta) - g_{i}(\btheta_k)| ] < \varepsilon/4
$$
for $ k \in 1, \ldots, K$.
Finally, the conclusion of the theorem follows from Kolmogorov's law of large numbers (KLLN). Taking $n \rightarrow \infty$, for any $k \in 1, \ldots, K$, 
$\sum_{i=1}^n g_i(\btheta_k)/n \stackrel{a.s.}{\rightarrow} 0$ since $\Expect[g_i(\btheta_k)]=0$. Moreover, $$ n^{-1}\sum_{i=1}^{n} \sup_{\btheta \in \Theta_k^K } |g_{i}(\btheta) - g_{i}(\btheta_k)| \stackrel{a.s.}{\rightarrow} \Expect[\sup_{\btheta \in \Theta_k^{K}}|g_{i}(\btheta) - g_{i}(\btheta_k)| ].$$
We can always choose $n>N(K, \kappa)$ for any small value $\kappa>0$ such that $\Pr[  |n^{-1}\sum_{i=1}^{n}g_{i}(\btheta_k)| > \varepsilon/2 ]< \kappa/(2K)$ and $\Pr[  n^{-1}\sum_{i=1}^{n} \sup_{\btheta \in \Theta_k^K } |g_{i}(\btheta) - g_{i}(\btheta_k)| > \varepsilon/2 ] < \kappa/(2K)$.
\end{proof}

We now divide the proof of Theorem~~\ref{thm:consistency} into three major steps. We consider the compact parameter space $\Theta$ with the following form:
$c_1 \le \sigma_1, \sigma_2, \sigma_3 \le c_2$, $\rho_1, \rho_2 \in [-1+\delta, 1 - \delta]$ for some $\delta>0$, $\vb^\top \vb \le c_3$, $ a_0, a_1, a_3 \in [-M, M]$, where $c_1$ and $\delta$ are small positive constants and $c_2, c_3, M$ are large positive constants. The true parameter $\btheta_0$ is assumed to be an interior point of $\Theta$ and the $\z$ is assumed to have the positive definite covariance matrix $\mSigma_z$.

\noindent \textbf{Step 1: Point-wise convergence in probability.}  
Let $\bar{L}_{\mN}(\btheta) = L_{\mN}(\btheta)/n$ and note that $\hat{\btheta}$ is the maximizer of $\bar{L}_{\mN}(\btheta)$. For any $\btheta = (\valpha, \Lambda), $ where $ \valpha =  (a_0, a_1,  \vb)$ and $\Lambda = ( \rho_1, \sigma_1, \sigma_2)$, we have
\begin{equation} \label{eqn:S1_obj}
    \begin{aligned}
    \bar{L}_{\mN} (\btheta)
     =  & \frac{n_1'}{n}\Big\{ -\log \det (\mSigma)  \\
     &\qquad - \sum_{i \in S_1} \frac{1}{n_1'}(y_i - \mu_{1i}, y_{n+i} - \mu_{3i})  \mSigma^{-1} (y_i - \mu_{1i}, y_{n+i} - \mu_{3i})^\top \Big\} \\
    & +  \frac{n_1 - n_1'}{n}\left\{- \log(\sigma_1^2) - \sum_{i \in C_1 \setminus S_1} \frac{1}{n_1 - n_1'}  \sigma_1^{-2}(y_i-\mu_{1i})^2\right\} \\
    & + \frac{n_2}{n}\left\{-\log(\sigma_2^2) - \sum_{i \in T_2} \frac{1}{n_2} \sigma_2^{-2}(y_i-\mu_{2i})^2 \right\}\\
   & \stackrel{p}{\rightarrow} \ell(\btheta) =  r_1 r_1'\left\{-\log \det (\mSigma) - \tr(\mV_1^{(1)} \mSigma^{-1})\right\}\\
   & + r_1 (1 - r_1') \{- \log(\sigma_1^2) - \sigma_1^{-2} V_2^{(1)} \} \\
   & + (1 - r_1) \{-\log(\sigma_2^2) - \sigma_2^{-2} V_3^{(1)}\} \\
   & = w_1 \ell_1(\btheta) + w_2 \ell_2(\btheta) + w_3 \ell_3(\btheta), 
\end{aligned}
\end{equation}
as $n_1' \rightarrow \infty, n_2 \rightarrow \infty$, where $w_1 = r_1r_1', w_2 = r_1(1-r_1')$ and $w_3 = (1-r_1)$.
The forms of $\mV_1^{(1)}, V_2^{(1)}$ and $ V_3^{(1)}$ are given by 
\begin{align*}
   &  \mV_1^{(1)} = \begin{bmatrix}
  q(\vb_0 - \vb; \mSigma_z) + \sigma_{10}^2 & q(\vb_0 - \vb; \mSigma_z) + \rho_{10}\sigma_{10}\sigma_{20}\\
 q(\vb_0 - \vb; \mSigma_z) + \rho_{10}\sigma_{10}\sigma_{20} & (a_{10} - a_1)^2 + q(\vb_0 - \vb; \mSigma_z) + \sigma_{20}^2
\end{bmatrix},  \\
& V_2^{(1)} = q(\vb_0 - \vb; \mSigma_z) + \sigma_{10}^2, \\
& V_3^{(1)} = (a_{00} - a_0)^2 + (a_{10} - a_1)^2 + 2(a_{00} - a_0)(a_{10} - a_1)+ q(\vb_0 - \vb; \mSigma_z) + \sigma_{20}^2,
\end{align*}
respectively, where $q(\vb_0 - \vb; \mSigma_z)$ represents the quadratic form $(\vb_0 - \vb)^\top \mSigma_z (\vb_0 - \vb)$.

\noindent \textbf{Step 2: Uniqueness of the maximizer.} The limit of the objective function $\ell(\btheta)$ enjoys the following decomposition 
\begin{equation}\label{eqn:obj_decomp}
    \ell(\btheta) = \sum_{k=1}^{3} w_k \ell_k(\btheta) = \sum_{k=1}^{3} w_k\Bigg[ f_k( \valpha, \Lambda ) + h_k( \Lambda) \Bigg],    
\end{equation}
where $f_k$ and $h_k$ are defined below.
Consider the case of $k = 1$, we have 
\begin{align*}
    f_1 = &  - \frac{1}{(1 - \rho^2)} \Big[\frac{(\vb_0 - \vb)\mSigma_z(\vb_0 - \vb)  }{\sigma_1^2} \\
    & - 2\rho \frac{(\vb_0 - \vb)^\top \mSigma_z (\vb_0 - \vb)}{\sigma_1 \sigma_2} + \frac{ (a_{10} - a_1)^2 + (\vb_0 - \vb)\mSigma_z (\vb_0 - \vb)}{\sigma_2^2} \Big] \\
    = & -\left( q(\vb_0 - \vb; \mSigma_z) + q( (0, a_{10} - a_1)^\top; \mI ) \right) \det(\mSigma^{-1}), \\
    f_2 = & -\frac{(\vb_0 - \vb)\mSigma_z(\vb_0 - \vb)  }{\sigma_1^2}, \\
    f_3 = & -\frac{(\vb_0 - \vb)\mSigma_z(\vb_0 - \vb)  }{\sigma_2^2} -\frac{(a_{00} + a_{10} - a_0 - a_1)^2}{\sigma_2^2},\\
    h_1  = &  -\log \det(\mSigma) - \tr(\mSigma^{-1} \mSigma_0),  \\
    h_2 = & -\log(\sigma_1^2) - \frac{\sigma_{10}^2}{\sigma_1^2} \\
    h_3 = & -\log(\sigma_2^2) - \frac{\sigma_{20}^2}{\sigma_2^2}
\end{align*}
where 
\begin{align*}
  \mSigma =  \begin{bmatrix}
\sigma_1^2 & \rho \sigma_1 \sigma_2 \\
\rho \sigma_1 \sigma_2 & \sigma_2^2
\end{bmatrix}, 
\qquad 
\quad \mSigma_0 =  \begin{bmatrix}
\sigma_{10}^2 & \rho_0 \sigma_{10} \sigma_{20} \\
\rho_0 \sigma_{10} \sigma_{20} & \sigma_{20}^2
\end{bmatrix}.
\end{align*}
 Observe that the $f_k$s are combination of quadratic forms and all of them attain the maximum value $0$ only if $\vb = \vb_0, a_0 = a_{00}, a_1 = a_{10}$ for any positive definite $\mSigma$. Regarding $h_k$'s, the strict concavity with respect to $\mSigma^{-1}$ implies that $\mSigma = \mSigma_0$ is the unique maximizer of $h_1$. Similarly, $\sigma_{1}=\sigma_{10}$ ($\sigma_2 = \sigma_{20}$) is the unique maximizer for $h_2$ ($h_3$). Hence, $\btheta_0$ is the unique maximizer of $\ell(\btheta)$.
 
 \noindent \textbf{Step 3: Uniform convergence in probability.}
Take $\vx_i = (y_i, \z_i)$ to be the pair of response and covariate. 
Let 
\begin{equation} \label{eqn:bivariate}
\begin{aligned}
    \ell_{1, \btheta}(\vx_i, \vx_{n+i})  = & -\log\det(\mSigma) \\
   & - (y_i -\mu_{1i}, y_{n+i} 
    - \mu_{3i}) \mSigma^{-1} (y_i -\mu_{1i}, y_{n+i} - \mu_{3i})^\top
\end{aligned}
\end{equation}
which satisfies  $\Expect[\ell_{1,\btheta}] = \ell_{1}(\btheta)$.
Then
\begin{align*}
   & \Expect[\sup_{\btheta \in \Theta} |\ell_{1,\btheta}(\vx_i, \vx_{n+i})|] \\
   \le & \Expect[ \sup_{\btheta \in \Theta}|\log \det (\mSigma )| + \sup_{\btheta \in \Theta} (y_i -\mu_{1i}, y_{n+i} - \mu_{3i}) \mSigma^{-1} (y_i -\mu_{1i}, y_{n+i} - \mu_{3i})^\top ] \\
    \le & \Expect[ \sup_{\btheta \in \Theta}|\log \det (\mSigma )| \\
    & \qquad + \sup\lambda_1(\mSigma^{-1}) \Expect[ \sup_{\btheta \in \Theta} (y_i -\mu_{1i}, y_{n+i} - \mu_{3i}) (y_i -\mu_{1i}, y_{n+i} - \mu_{3i})^\top ] \\
    \stackrel{(i)}{\le} & M_1 
    + \sup \lambda_1(\mSigma^{-1})  \sup \{\sigma_{10}^2 + \sigma_{20}^2 +  2\|\vb_0 - \vb\|^2\tr{\mSigma_z} + (a_{10} - a_1)^2 \} \\
    \stackrel{(ii)}{\le} & M_1 + M_2.
\end{align*}
In the above derivations, (i) is due to the fact that quadratic form $\vv^\top \mSigma^{-1}\vv \le \lambda_1(\mSigma^{-1})\|\vv\|^2 $ for any vector $\vv$ and $\|\vz_i^\top (\vb- \vb_0)\|^2  \le \|\vz_i\|^2 \|\vb - \vb_0\|^2$; (ii) is because of $\btheta = (a_0, a_1, \vb, \rho_1, \sigma_1, \sigma_2) \in [-M, M]^2 \times \mathbb{B}_{p}(c_3) \times [-1+\delta, 1-\delta] \times [c_1, c_2] \times [c_1, c_2] $. Define $g_i(\btheta) := \ell_{1,\btheta}( (\vx_i, \vx_{n+i}) ) - \Expect[\ell_{1,\btheta} ( (\vx_i, \vx_{n+i}))]$. Then the uniform convergence of $n_1'^{-1} \sum_{i=1}^{n_1'} \ell_{1,\btheta}$ to $\ell_{1}(\btheta)$ follows from Lemma~\ref{lem:uniform}. In the same spirit, we can also define 
\begin{equation}
    \begin{split}
       & \ell_{2,\btheta}(\vx_i) = - \log(\sigma_1^2) -  (y_i-\mu_{1i})^2/\sigma_1^2, \\
        & \ell_{3,\btheta}(\vx_i) = -\log(\sigma_2^2) - (y_i-\mu_{2i})^2/ \sigma_2^{2}.
    \end{split}
\end{equation}
We shall have similar conclusion regarding $\ell_{2, \btheta}(\vx_i)$ and $ \ell_{3,\btheta}(\vx_i)$.
The uniform convergence in probability of $\bar{L}_{\mN}(\btheta)$ to $\ell(\btheta)$ is then verified. Finally, we employ Lemma~\ref{lem:takeshi} to establish the consistency of $\hat{\btheta}$.

\subsection{Proof of Theorem~\ref{thm: asymptotic}} \label{apx:subsec:asymptotic}
Follow the notation used in \cite{vaart_1998},
we denote 
\begin{align*}
   & \bbP_{n_1'}^{(1)} \ell_{1,\btheta} = (n_1')^{-1} \sum_{i=1}^{n_1'} \ell_{1,\btheta}(\vx_i, \vx_{n+i}), & i = 1, \ldots, n_1', \\
   & \bbP_{n_1 - n_1'}^{(2)}  \ell_{2,\btheta} = (n_1 - n_1')^{-1} \sum_{i=1}^{n_1 - n_1'} \ell_{2,\btheta}(\vx_i), & i = n_1'+1, \ldots, n_1, \\
   & \bbP_{n_2}^{(3)} \ell_{3, \btheta} = n_2^{-1} \sum_{i=1}^{n_2} \ell_{3, \btheta}(\vx_i), & i = n_1, \ldots, n.
\end{align*}
The objective function $\bar{L}_{\mN}(\btheta)$ can be written as 
$$
\bar{L}_{\mN}(\btheta) = \frac{n_1'}{n} \bbP_{n_1'}^{(1)}  \ell_{1, \btheta} +\frac{n_1 - n_1'}{n}  \bbP_{n_1 - n_1'}^{(2)} \ell_{2, \btheta} + \frac{n_2}{n} \bbP_{n_2}^{(3)} \ell_{3,\btheta}.
$$
Our (multivariate) Gaussian log-likelihood satisfies the Lipschitz-Hessian condition~\citep{carmon2018accelerated,nesterov2006cubic} that
$$ \| \nabla^2 \ell_{1, \btheta_1}(\vx_i, \vx_{n+i}) -  \nabla^2 \ell_{1, \btheta_2 }(\vx_i, \vx_{n+i}) \|_{op} \le M_1 (\vx_i, \vx_{n+i}) \|\btheta_1 - \btheta_2 \|, $$
and $\| \nabla^2 \ell_{k, \btheta_1}(\vx_i) -  \nabla^2 \ell_{k, \btheta_2 }(\vx_i) \|_{op} \le M_k(\vx_i)\|\btheta_1 - \btheta_2 \|, k = 2,3$ for some absolutely integrable functions $M_1 (\vx_i, \vx_{n+i}), M_2(\vx_i)$ and $M_3(\vx_i)$ in the sense that $\Expect[ M_{k}] < \infty, k = 1,2,3$, as we will verify later in Section~\ref{apx:sec:Lipschitz-Hessian}.
By the first-order condition for $\hat{\btheta}$, we have
\begin{equation*}
    \begin{aligned}
     0 = & \nabla \bar{L}_{\mN}(\hat{\btheta}) = \frac{n_1'}{n} \bbP_{n_1'}^{(1)} \nabla \ell_{1, \btheta_0} +\frac{n_1 - n_1'}{n}  \bbP_{n_1 - n_1'}^{(2)} \nabla \ell_{2, \btheta_0} + \frac{n_2}{n} \bbP_{n_2}^{(3)} \nabla \ell_{3,\btheta_0} \\
     & + \Big(\frac{n_1'}{n} \bbP_{n_1'}^{(1)} \nabla^2 \ell_{1, \btheta_0} + \frac{n_1 - n_1'}{n}  \bbP_{n_1 - n_1'}^{(2)} \nabla^2 \ell_{2, \btheta_0} + \frac{n_2}{n}  \bbP_{n_2}^{(3)} \nabla^2 \ell_{3,\btheta_0}\Big)(\hat{\btheta} - \btheta_0) \\
     & + \hat{\gamma} (\hat{\btheta} - \btheta_0), 
\end{aligned}
\end{equation*}
where $\hat{\gamma} =
    \int_{0}^1 \big( \nabla^2  \bar{L}_{\mN}((1-t)\hat{\btheta} + t \btheta_0) - \nabla^2 \bar{L}_{\mN}(\btheta_0) \big) \intd t
$
and 
\begin{equation}\label{eqn:residual}
    \begin{aligned}
   \|\hat{\gamma} (\hat{\btheta} - \btheta_0)\| & \le \int_{0}^1 \| \big( \nabla^2 \bar{L}_{\mN}((1-t)\hat{\btheta} + t \btheta_0) - \nabla^2 \bar{L}_{\mN}(\btheta_0) \big) \|_{op} \|\hat{\btheta} - \btheta_0\| \intd t \\
    & \le  \Big[ \frac{n_1'}{n_1} \bbP_{n_1'}^{(1)} M_1 + \frac{n_1 - n_1'}{n} \bbP_{n_1 - n_1'}^{(2)} M_2  + \frac{n_2}{n}\bbP_{n_2}^{(3)}  M_3 \Big] \frac{\|\hat{\btheta} - \btheta_0\|^2}{2}    \\
   & = O_p(\|\hat{\btheta} - \btheta_0)\|^2).
\end{aligned}
\end{equation}
Re-arranging the terms, we have 
\begin{align*}
   & \sqrt{n}(\hat{\btheta} - \btheta_0)  =\\
   & - \Bigg(\frac{n_1'}{n} \bbP_{n_1'}^{(1)} \nabla^2 \ell_{1, \btheta_0} + \frac{n_1 - n_1'}{n}  \bbP_{n_1 - n_1'}^{(2)} \nabla^2 \ell_{2, \btheta_0} + \frac{n_2}{n}  \bbP_{n_2}^{(3)} \nabla^2 \ell_{3,\btheta_0} + o_p(1) \Bigg)^{-1} \\
   & \Bigg( \frac{n_1'}{\sqrt{n}} \bbP_{n_1'}^{(1)} \nabla \ell_{1, \btheta_0} +\frac{n_1 - n_1'}{\sqrt{n}}  \bbP_{n_1 - n_1'}^{(2)} \nabla \ell_{2, \btheta_0} + \frac{n_2}{\sqrt{n}} \bbP_{n_2}^{(3)} \nabla \ell_{3,\btheta_0} \Bigg).
\end{align*}
Given the facts that $n_1/n \rightarrow w_1 = r_1$, $n_1'/n_1 \rightarrow w_2 = r_1', (n_1 - n_1')/n_1 \rightarrow w_3 = 1-r_1'$, $\Expect \dot{\ell}_{k,\btheta_0} = 0 $ and the sample points from case and control groups of different batches are independent, the Linderberg-Feller conditions can be verified as follows. Define
\begin{equation*}
\begin{cases}
    X_{n, i} = (\dot{\ell}_ {1, \btheta}(\vx_i, \vx_{n+i})/\sqrt{n}, 0, 0)^\top, & i = 1, \ldots, n_1', \\
    X_{n, i} = (0,\dot{\ell}_ {2, \btheta}(\vx_i)/\sqrt{n}, 0)^\top, & i = n_1'+1, \ldots, n_1, \\
    X_{n,i} =  (0, 0, \dot{\ell}_ {3, \btheta}(\vx_i)/\sqrt{n})^\top, & i = n_1, \ldots, n. 
\end{cases}
\end{equation*}
Then for any $\epsilon > 0$, 
\begin{align*}
    & \sum_{i=1}^{n} \Expect\Big[ \|X_{n,i}\|^2\v1 \{\|X_{n,i}\| > \epsilon \} \Big]  \\
    & \le \max_{k} \Expect\Big[|\dot{\ell}_{k,\btheta}|^2 \v1\{ |\dot{\ell}_{k,\btheta}/\sqrt{n}| > \epsilon \} \Big].
\end{align*}
Since $\Expect[|\dot{\ell}_{k,\btheta}|^2] < \infty$ for 
$k = 1,2,3$, it suffices to show that 
$$
\v1\{ |\dot{\ell}_{k,\btheta}/\sqrt{n}| > \epsilon \} \rightarrow^{a.s} 0,
$$
which is true because $1/\sqrt{n} \rightarrow 0$.
 Meanwhile,
\begin{equation} \label{eqn:asymp_Cov}
    \begin{aligned}
   & \sum_{i=1}^n \cov(X_{n,i}) \rightarrow \\ & \begin{pmatrix}
      r_1 r_1' \Expect[\dot{\ell}_{1,\btheta_0}\dot{\ell}_{1,\btheta_0}^\top] & & \\
       & r_1(1-r_1')\Expect[\dot{\ell}_{2,\btheta_0}\dot{\ell}_{2,\btheta_0}^\top] & \\
       & & (1-r_1)\Expect[\dot{\ell}_{3,\btheta_0}\dot{\ell}_{3,\btheta_0}^\top]
    \end{pmatrix}.
\end{aligned}
\end{equation}
Thus, $
    \sum_{i=1}^n X_{n,i} = (\frac{n_1'}{\sqrt{n}}\bbP_{n_1'}^{(1)}\dot{\ell}_{1,\btheta_0}, \frac{n_1 - n_1'}{\sqrt{n}}\bbP_{n_1 - n_1'}^{(2)}\dot{\ell}_{2,\btheta_0}, \frac{n_2}{\sqrt{n}}\bbP_{n_2}^{(3)}\dot{\ell}_{3,\btheta_0} )^\top 
$
is jointly normal with the asymptotic covariance in~\eqref{eqn:asymp_Cov}. The conclusion thus follows.

\subsubsection{Verification of the Lipschitz-Hessian Condition}\label{apx:sec:Lipschitz-Hessian}
We now verify the Lipschitz-Hessian condition under our model setting. For any positive definite symmetric matrix $\mA$, we have 
$
\|\mA\|_{op}\le \|\mA\|_{F}
$, where $\|\mA\|_{F}$ denotes the Frobenius norm of $\mA$. For $\btheta=(\theta_1,\dots,\theta_q)\in\mathbb{R}^q$, we have 
\begin{align*}
    \| \nabla^2 \ell_{1, \btheta_1} - \nabla^2 \ell_{1, \btheta_2} \|_{op} & \le \| \nabla^2 \ell_{1, \btheta_1} - \nabla^2 \ell_{1, \btheta_2} \|_{F} \\
    & = \Big( \sum_{s=1}^q \sum_{t=1}^q |(\nabla^2 \ell_{1, \btheta_1} - \nabla^2 \ell_{1, \btheta_2})_{st}|^2 \Big)^{1/2}, 
\end{align*}
and for any $s, t \in \{1, \ldots, q\}$,
\begin{align*}
    \Big|\frac{\partial^2 \ell_{1, \btheta_1}(\vx_i, \vx_{n+i}) }{\partial \theta_s \partial \theta_t } - \frac{\partial^2 \ell_{1, \btheta_2} (\vx_i, \vx_{n+i}) }{\partial \theta_s \partial \theta_t }\Big| \le M_{st}(\vx_i, \vx_{n+i}) \|\btheta_1 - \btheta_2\|.
\end{align*}
with $M_{st}(\vx_i, \vx_{n+i}): = \sup_{\btheta}  \|\nabla \frac{\partial^2 \ell_{1, \btheta}(\vx_i, \vx_{n+i}) }{\partial \theta_s \partial \theta_t }\|$ because of the smoothness of the log-likelihood and the compactness of $\Theta$. Hence, 
$$\| \nabla^2 \ell_{1, \btheta_1} - \nabla^2 \ell_{1, \btheta_2} \|_{op} \le \left(\sum_{s=1}^q \sum_{t=1}^q M_{st}^2(\vx_i, \vx_{n+i})\right)^{1/2} \|\btheta_1 - \btheta_2 \|.$$
Using the form of $\ell_{1,\btheta}$, we have the following observations:
\begin{itemize}
    \item Taking any derivative w.r.t the variance/covariance parameters ($\rho, \sigma_1$ and $\sigma_2$) will not change the degree of the polynomials (w.r.t to $y_i$'s). For instance,
    \begin{align*}
    \frac{\partial^3 \ell_{1,\btheta}}{\partial \sigma_1^3} = -\frac{4}{\sigma_1^3} + \frac{1}{1-\rho^2}\Big[ \frac{12(y_i - \z_i^\top \vb)^2 }{\sigma_1^5} + \frac{12\rho (y_i - \z_i^\top \vb) (y_{n+i} - \z_i^\top \vb - a_1)}{\sigma_1^4 \sigma_2}\Big].
    \end{align*}
    \item Any third derivative w.r.t $a_0, a_1$ and $\vb$ is $0$. This can be seen from 
    \begin{align*}
    \frac{\partial^3 \ell_{1,\btheta}}{ \partial a_0^3 } = 0, \quad
    \frac{\partial^3 \ell_{1,\btheta}}{ \partial a_1^3 } = 0, \quad \frac{\partial^2 \ell_{1,\btheta}}{\partial \vb^\top \partial \vb} = \det(\mSigma) \vz_i \vz_i^\top.
    \end{align*}
    Moreover, the third derivative w.r.t any component $b_j$ in $\vb$ is zero. 
\end{itemize}
One can verify that all the third derivatives are dominated by some terms that are proportional to $\vz_i\vz_i^\top$, $(\epsilon_i^{(1)})^2$ and $(\epsilon_{n+i}^{(2)})^2$. 
In particular, we have 
$$
M_1(\vx_i, \vx_{n+i}) = A(\v1^\top \z_i \vz_i^\top \v1) + B(\epsilon_i^{(1)})^2 + C(\epsilon_{n+i}^{(2)})^2,
$$
for sufficiently large positive constants $A, B$ and $C$ such that
$$\| \nabla^2 \ell_{1, \btheta_1} - \nabla^2 \ell_{1, \btheta_2} \|_{op} \le M_{1}(\vx_i, \vx_{n+i}) \|\btheta_1 - \btheta_2 \|.$$
We have assumed that $\Expect[\vz_i\vz_i^\top] = \mSigma_z$, $\Var[\epsilon_i^{(1)}] = \sigma_1^2$ and $\Var[\epsilon_i^{(2)}] = \sigma_2^2$ in Condition~\ref{ass: compact}-\ref{ass:portion}, which implies that $\Expect[M_1] < \infty$. Similar arguments apply to $\ell_{2,\btheta}$ and $\ell_{3,\btheta}$.

\section{Computational algorithm and statistical inference}\label{apx:sec:updating}
In this section, we describe the parameters updating scheme in detail. We first introduce some notation. Let $R_1= \sum_{i \in S_1} (y_i - \z_i^\top \b )/n_1'$, $R_2 = \sum_{i \in T_2}( y_i - \z_i^\top \b)/n_2$, $R_3 = \sum_{i \in S_1}(y_{n+i} - \z_{n+i}^\top \b)/n_1'$. Let $W_{(S_1)} = \sum_{i \in S_1} (y_i - \mu_{1i})^2$,  $W_{(C_2)} = \sum_{i \in S_1} (y_{n+i} - \mu_{3i})^2$, $W_{(S_1 \cdot C_2)} = \sum_{i \in S_1} (y_i - \mu_{1i}) (y_{n+i} - \mu_{3i}) $, $ W_{(C_1 \setminus S_1)} = \sum_{i \in C_1 \setminus S_1}(y_i - \mu_{1i})^2$, and $W_{(T_2)} = \sum_{i \in T_2} (y_i - \mu_{2i})^2$.

Taking the first-order derivative of the objective function with respect to $a_1$ and $a_0$ separately and setting the expressions to $0$, we obtain
\begin{align*}
         & \frac{n_1' (R_3 - a_1) }{\sigma_2^2 (1 - \rho^2)} - \frac{\rho n_1' R_1}{\sigma_1 \sigma_2 (1 - \rho^2)} + \frac{n_2(R_2 - a_0 - a_1)}{\sigma_2^2}  = 0, \\
         & R_2 - a_0 - a_1  = 0.
\end{align*}
The explicit forms of the updating rules for $a_0$ and $a_1$ are given respectively by
\begin{equation}
    a_0 = R_2 - (R_3 - \frac{\rho\sigma_2}{\sigma_1}R_1), \label{eqn:a0_S1}
\end{equation} 
    and 
\begin{equation}\label{eqn:a1_S1}
        a_1 = R_3 - \frac{\rho\sigma_2}{\sigma_1}R_1.
\end{equation}
For $\rho, \sigma_1$ and $\sigma_2$, there is no closed-form updating rule. The correlation $\rho$ is updated by finding the real positive root of the cubic equation
$$
    n_1' \rho(1-\rho^2) = \rho(\frac{W_{(S_1)}}{\sigma_1^2}+ \frac{W_{(C_2)}}{\sigma_2^2}) - \frac{(1+\rho^2)W_{(S_1\cdot C_2)}}{\sigma_1 \sigma_2}.
$$
The standard deviations $\sigma_1$ and $\sigma_2$ are updated via finding the positive roots of the following two quadratic equations
\begin{equation*}
        \begin{split}
           &   n_1 (1-\rho^2) \sigma_1^2 = W_{(S_1)} + (1-\rho^2) W_{(C_1\setminus S_1)}  - \frac{\rho W_{(S_1\cdot C_2)} \sigma_1}{\sigma_2},\\
          & (n_1' + n_2 )(1 - \rho^2)\sigma_2^2 = W_{(C_2)} + (1- \rho^2)W_{(T_2)} - \frac{\rho W_{(S_1\cdot C_2)} \sigma_2}{\sigma_1}.
        \end{split}
\end{equation*}

\begin{algorithm2e}[t]
	\KwIn{initial value $\theta^{(0)} = (a_0^{(0)}, a_1^{(0)}, \b^{(0)}, \rho^{(0)}, \sigma_1^{(0)}, \sigma_2^{(0)} )$}
	\KwOut{ MLE $\hat{\btheta} = (\hat{a}_0, \hat{a}_1, \hat{\b}, \hat{\rho}, \hat{\sigma}_1, \hat{\sigma}_2)$}
	\BlankLine
	\DontPrintSemicolon
	Set $k = 1$ \;
    Set $\mu_{1i}^{(0)} = \vz_i^\top \vb^{(0)}$, $\mu_{2i}^{(0)} = a_0^{(0)} + a_1^{(0)} + \vz_i^\top \vb^{(0)}$, $\mu_{3i}^{(0)} = a_1^{(0)} + \vz_i^\top \vb^{(0)}$. Denote $W_{(S_1)}^{(0)} = \sum_{i \in S_1} (y_i - \mu_{1i}^{(0)})^2$,  $W_{(C_2)}^{(0)} = \sum_{i \in S_1} (y_{n+i} - \mu_{3i}^{(0)})^2$, $W_{(S_1\cdot C_2)}^{(0)} = \sum_{i \in S_1} (y_i - \mu_{1i}^{(0)}) (y_{n+i} - \mu_{3i}^{(0)}) $, $ W_{(C_1 \setminus S_1)}^{(0)} = (y_i - \mu_{1i}^{(0)})^2,  W_{(T_2)}^{(0)} = \sum_{i \in T_2} (y_i - \mu_{2i}^{(0)})^2$. \;
	\Repeat{convergence.}{
		Compute $\rho^{(k)}$ by solving
        $n_1' \rho(1-\rho^2) = \rho \Big(\frac{W_{(S_1)}^{(k-1)}}{ (\sigma_1^{(k-1)})^2 }+ \frac{W_{(C_2)}^{(k-1)}}{(\sigma_2^{(k-1)})^2} \Big) - \frac{(1+\rho^2)W_{(S_1\cdot C_2)}^{(k-1)}}{\sigma_1^{(k-1)} \sigma_2^{(k-1)}}.
        $ \; 
		Compute $\sigma_1^{(k)}$ by solving 
		$
        n_1 \big(1-(\rho^{(k)})^2 \big) \sigma_1^2 = W_{(S_1)}^{(k-1)} + \big(1-(\rho^{(k)})^2 \big) W_{(C_1\setminus S_1) }^{(k-1)}  - \frac{\rho^{(k)} W_{(S_1\cdot C_2)}^{(k-1)} \sigma_1}{\sigma_2^{(k-1)}}.
        $ \; 
		Compute $\sigma_2^{(k)}$ by solving $$
        (n_1' + n_2 )\big(1-(\rho^{(k)})^2 \big)\sigma_2^2 = W_{(C_2)}^{(k-1)} + \big(1-(\rho^{(k)})^2 \big) W_{(T_2)}^{(k-1)} - \frac{\rho^{(k)} W_{(S_1\cdot C_2)}^{(k-1)} \sigma_2}{\sigma_1^{(k-1)}}.
        $$ \;
		Compute $a_1^{(k)} = R_3^{(k-1)} - \frac{\rho^{(k)}\sigma_2^{(k)}}{\sigma_1^{(k)}}R_1^{(k-1)}$, $a_0^{(k)} = R_2^{(k-1)} - a_1^{(k)}$.  \;
		Update $\mS^{(k)}, \vt^{(k)}$ based on ~\eqref{matrix:S}, \eqref{matrix:t} given $\rho^{(k)}, \sigma_1^{(k)}, \sigma_2^{(k)}$. \;
		Compute $\vb^{(k)} = \mS^{(k)-1} \vt^{(k)} $ \; 
		Update $R_1^{(k)},R_2^{(k)},  R_3^{(k)}$ given $\vb^{(k)}$ \;
		Update $W_{(S_1)}^{(k)}, W_{(C_2)}^{(k)}, W_{(S_1\cdot C_2)}^{(k)}, W_{(C_1\setminus S_1)}^{(k)}, W_{(T_2)}^{(k)}$ given $a_{0}^{(k)}, a_1^{(k)}, \vb^{(k)}$ \; 
		Set $k=k+1$. \;
	}
	The final MLE estimator is $\hat{\btheta} = (\hat{a}_0, \hat{a}_1, \hat{\b}, \hat{\rho}, \hat{\sigma}_1, \hat{\sigma}_2) = (a_0^{(k)}, a_1^{(k)}, \b^{(k)}, \rho^{(k)}, \sigma_1^{(k)}, \sigma_2^{(k)} )$.
	\caption{The alternate parameter updating for MLE. \label{algorithm:S1}}
\end{algorithm2e}
To describe the updating rule for $\vb$, we need to introduce some additional symbols. Suppose the dimension of $\vb$ is $p$. Write $\mZ^{(S_1)} = \mZ^{(C_2)} =  (\vz_1, \ldots, \vz_{n_1'})^\top$ as the $n_1' \times p$ matrix of remeasured covariates with the rows being the covariates for each sample. Let $\vy^{(S_1)} = (y_1, \ldots, y_{n_1'})^\top$ be the corresponding response vector of the control group in the first batch. Also, we define $\mZ^{(T_2)}$ as the $n_2 \times p$ design matrix of the treatment group in the second batch and $\vy^{(T_2)}$ as the corresponding response vector. Let $\mZ^{(C_1\setminus S_1)} = (\vz_{n_1'+1}, \ldots, \vz_{n_1})$ be the $(n_1 - n_1') \times p$ matrix of covariates that are not remeasured and 
$\vy^{(C_1\setminus S_1)}$ be the corresponding response in the first batch. Moreover, we let $\vy^{(C_2)} = (y_{n+1}, \ldots, y_{n + n_1'})^\top$ be the vector of responses of the remeasured samples in the control group in the second batch. For any $n \times p$ matrix $\mZ$, we define $\bar{\mZ}=\mZ^\top\mathbf{1}/n$ as the $p \times 1$ vector that contains the mean value for each column of $\mZ$, and let $\mZ_c$ be the centralized matrix by subtracting the mean vector $\bar{\mZ}$ from each row of $\mZ$. Similarly, we let $\vy_c$ be the centralized vector by subtracting the mean value from each element in $\vy$.

Using the first order condition for $\vb$, and \eqref{eqn:a0_S1} and \eqref{eqn:a1_S1}, we have 
\begin{equation}\label{eqn:solveb}
    \begin{aligned}
   & \frac{\mZ^{ (C_2) \top} (\vy^{(S_1)} - \mZ^{(C_2) }\vb)}{\sigma_1^2 (1 - \rho^2)} - \frac{\rho \mZ^{(C_2) \top}(\vy^{(C_2)} - \mZ^{(C_2)}\vb +R_3 - \frac{\rho\sigma_2}{\sigma_1}R_1 ) }{\sigma_1 \sigma_2(1 - \rho^2) } \\
   & - \frac{\rho (\mZ^{(C_2)} - \frac{\rho \sigma_2}{\sigma_1}\bar{\mZ}^{(C_2)})^\top (\vy^{(S_1)} - \mZ^{(C_2)}\vb ) }{\sigma_1 \sigma_2 (1 - \rho^2)} \\
   & + \frac{(\mZ^{(C_2)} - \frac{\rho \sigma_2}{\sigma_1} \bar{\mZ}^{(C_2)})^\top (\vy^{(C_2)} - \mZ^{(C_2)}\vb +R_3 - \frac{\rho\sigma_2}{\sigma_1}R_1 ) }{\sigma_2^2 (1 - \rho^2)}\\
   & + \frac{\mZ^{ (C_1 \setminus S_1) \top} (\vy^{ (C_1 \setminus S_1 )} - \mZ^{(C_1 \setminus S_1) } \vb)}{\sigma_1^2}+ \frac{\mZ^{ (T_2) \top}_c (\vy^{(T_2)}_c - \mZ^{(T_2)}_c \vb)}{\sigma_2^2} = 0.
\end{aligned}
\end{equation}
Re-arranging the above equation by putting the terms related to $\vb$ on the left-hand side and the rest on the right-hand side,
we obtain a linear equation $\mS \b= \vt$, where $\mS$ and $\vt$ depend on $(\rho, \sigma_1, \sigma_2)$.
The forms of $\mS$ and $\vt$ are given respectively by
\begin{equation}
    \label{matrix:S}
    \begin{aligned}
        & \mS  = \frac{\mZ^{(C_2)\top} \mZ^{(C_2)}}{\sigma_1^2 (1 - \rho^2)} - \frac{\rho \mZ^{(C_2) \top} (\mZ^{(C_2)} - (1 - \frac{\rho\sigma_2}{\sigma_1} )\bar{\mZ}^{(C_2)}) }{\sigma_1 \sigma_2 (1 - \rho^2)} \\
        & - \frac{\rho (\mZ^{(C_2)} - (1 - \frac{\rho\sigma_2}{\sigma_1} )\bar{\mZ}^{(C_2)})^\top \mZ^{(C_2)}}{\sigma_1 \sigma_2(1-\rho^2)} \\
        & +  \frac{(\mZ^{(C_2)} - (1 - \frac{\rho\sigma_2}{\sigma_1})\bar{\mZ}^{(C_2)})^\top(\mZ^{(C_2)} - (1 - \frac{\rho\sigma_2}{\sigma_1})\bar{\mZ}^{(C_2)})}{\sigma_2(1-\rho^2)}  \\
        & + \frac{\mZ^{(C_1 \setminus S_1)\top} \mZ^{(C_1 \setminus S_1) }}{\sigma_1^2} + \frac{\mZ_c^{(T_2)\top} \mZ_c^{(T_2)}}{\sigma_2^2},
    \end{aligned}
\end{equation}
and 
\begin{equation}
    \label{matrix:t}
    \begin{aligned} 
    \vt & = \frac{\mZ^{(C_2)\top} \vy^{S_1}}{\sigma_1^2 (1 - \rho^2)} - \frac{\rho \mZ^{(C_2) \top} (\vy^{(C_2)} - \bar{\vy}^{(C_2)} + \frac{\rho\sigma_2}{\sigma_1}\bar{\vy}^{(S_1)} ) }{\sigma_1 \sigma_2 (1 - \rho^2)} \\
        & - \frac{\rho (\mZ^{(C_2)} - (1 - \frac{\rho\sigma_2}{\sigma_1})\bar{\mZ}^{(C_2)})^\top \vy^{(S_1)}}{\sigma_1 \sigma_2(1-\rho^2)} \\
        & + \frac{(\mZ^{(C_2)} - (1 - \frac{\rho\sigma_2}{\sigma_1})\bar{\mZ}^{(C_2)})^\top(\vy^{(C_2)} - \bar{\vy}^{(C_2)} + \frac{\rho\sigma_2}{\sigma_1}\bar{\vy}^{(S_1)})}{\sigma_2^2(1-\rho^2)}  \\
        & + \frac{\mZ^{( C_1 \setminus S_1)\top} \vy^{ (C_1 \setminus S_1) }}{\sigma_1^2} + \frac{\mZ_c^{(T_2)\top} \vy_c^{(T_2)}}{\sigma_2^2}.
\end{aligned}
\end{equation}
We summarize the iterative updating procedure in Algorithm~\ref{algorithm:S1}.

To estimate the variance of $\hat{a}_0$, let $\tilde{\mZ} = (1 - \hat{\rho}\hsigma_2/\hsigma_1)\mZ^{(C_2)}$. Note that $\hat{a}_0$ can be written as
\begin{equation}\label{eqn:ReMeasure_deriv}
\begin{split}
           \hat{a}_0 = &  \frac{\v1_{n_2}^\top }{n_2}(\vy^{(T_2)} - \mZ^{(T_2)} \hat{\vb} ) - \hat{a}_1\\
   = & (-\frac{\v1_{n_2}^\top }{n_2} \mZ^{(T_2)}\mA + \frac{\v1_{n_1'}^\top}{n_1'} \tilde{\mZ}\mA + \frac{\v1_{n_1'}^\top}{n_1'}\frac{\hat{\rho} \hat{\sigma}_2}{\hat{\sigma}_1} )\vy^{(S_1)} + (\frac{\v1_{n_2}^\top }{n_2} - \frac{\v1_{n_2}^\top }{n_2} \mZ^{(T_2)} \mB + \frac{\v1_{n_1'}^\top}{n_1'} \tilde{\mZ} \mB) \vy^{(T_2)} \\
   + & (-\frac{\v1_{n_2}^\top }{n_2}\mZ^{(T_2)}\mC - \frac{\v1_{n_1'}^\top}{n_1'} + \frac{\v1_{n_1'}^\top}{n_1'} \tilde{\mZ} \mC) \vy^{(C_2)} + (-\frac{\v1_{n_2}^\top }{n_2}\mZ^{(T_2)} \mD + \frac{\v1_{n_1'}^\top}{n_1'}\tilde{\mZ}\mD ) \vy^{(C_1\setminus S_1)} \\
   = &  \vc_1^\top \vy^{(S_1)} + \vc_2^\top \vy^{(T_2)} +  \vc_3^\top \vy^{(C_2)} +  \vc_4^\top \vy^{(C_1\setminus S_1)},
\end{split}
\end{equation}
where $\mA, \mB, \mC$ and $\mD$ are the coefficient matrices such that $\hat{\vb} = \mA \vy^{(S_1)} + \mB \vy^{(T_2)} + \mC \vy^{(C_2)} + \mD \vy^{(C_1 \setminus S_1)}$. The explicit forms are given by
\begin{align*}
   & \mA = \hat{\mS}^{-1} \mZ^{(S_1)\top} \big( \frac{\mI_{n_1'}}{\hsigma_1^{2}(1 - \hrho^{2})} - \frac{\hrho}{\hsigma_1 \hsigma_2 (1 - \hrho^{2})} (\mI_{n_1'} - \frac{\v1_{n_1'}\v1_{n_1'}^\top}{n_1'}) - \frac{\hrho^{2}}{\hsigma_1^{2} (1 - \hrho^{2} ) } \frac{\v1_{n_1'} \v1_{n_1'}^\top}{n_1'} \big), \\
   & \mB =  \hat{\mS}^{-1} \frac{\mZ^{(T_2)\top}}{\hsigma_2^{2}} (\mI_{n_2} - \v1_{n_2}\v1_{n_2}^\top/n_2), \\
    &    \mC = \hat{\mS}^{-1} ( \frac{1}{\hsigma_2^{2}(1 - \hrho^{2})} - \frac{\hrho}{\hsigma_1 \hsigma_2(1 - \hrho^{2})} )\mZ^{(S_1)\top}(\mI_{n_1'} - \v1_{n_1'}\v1_{n_1'}^\top/n_1'), \\
     & \mD = \hat{\mS}^{-1} \frac{\mZ^{(C_1 \setminus S_1)\top }}{\hsigma_1^2 },
\end{align*}
where $\hat{\mS}$ is defined in the same way as $\mS$ by replacing $(\rho, \sigma_1, \sigma_2)$ with $(\hat{\rho}, \hat{\sigma_1}, \hsigma_2)$.
The variance of $\hat{a}_0$ can then be estimated by 
\begin{align}\label{eq-var}
    \widehat{\Var}(\hat{a}_0) = \hat{\sigma}_1^2 (\vc_1^\top \vc_1 + \vc_4^\top \vc_4) + \hat{\sigma}_2^2 (\vc_2^\top \vc_2 + \vc_3^\top \vc_3) + 2\hat{\rho}\hat{\sigma}_1 \hat{\sigma}_2 \vc_1^\top \vc_3.  
\end{align}

\section{The location-scale matching approach}\label{apx:sec:LS}
The location-scale~(LS) approach assumes a model for the location~(mean) and scale~(variance) with the batches. By standardizing the means and variances across the batches, the batch effect can then be removed. The estimation of the scale proportion is
$\hat{\sigma}_2/\hat{\sigma}_1$, where $\hat{\sigma}_2$ is the standard deviation of $\vy^{(C_2)}$, and $\hat{\sigma}_1$ is the standard deviation of $\vy^{(S_1)}$.
Then the batch-adjusted data for the first batch, $\vy_*^{(C_1)}$, are given by
\begin{align*}
   \vy^{(C_1)}_* =  \frac{\hat{\sigma}_2}{\hat{\sigma}_1}(\vy^{(C_1)} - \bar{\vy}^{(C_1)}) + \bar{\vy}^{(C_1)} + \bar{\vy}^{(C_2)} - \bar{\vy}^{(S_1)}.
\end{align*}
We can assume the adjusted control samples in the first batch, together with the case samples in the second batch $(\vy_*^{(C_1)}, \vy^{(T_2)} )$ follow the model
\begin{align*}
&\text{Control (batch 2):} \quad y_i = a_{1}+\z_i^\top \b + \epsilon_i^{(2)},\quad i=1,\dots,n_1, \\
&\text{Case (batch 2):} \quad y_i = a_{0}+a_{1} + \z_i^\top \b + \epsilon_i^{(2)},\quad i=n_1+1,\dots,n.
\end{align*}
Therefore, we can use the least squares to get the parameter estimates.

\section{Additional simulations}\label{apx:sec:add_simu}
\begin{table}[ht!] 
    \centering
    \resizebox{0.9\textwidth}{!}{
    \begin{tabular}{c| c c c c  }
    \hline 
    \hline
       & Batch2 & Ignore & LS & ReMeasure  \\
    \hline
    $a_1 = 0.5$  & \multicolumn{4}{|c}{$a_0 = 0.5$ } \\
\hline
    \hline
     \multicolumn{2}{c}{$ \sigma_1 = 0.5, \rho = 0.3$}    \\
\hline
$n_1' = 5$ & 0.209(0.010) & 0.267(0.005) & \textbf{0.206(0.010)} & 0.239(0.012)\\ 
$n_1' = 10$ & \textbf{0.111(0.005)} & 0.267(0.005) & 0.114(0.005) & 0.111(0.005)\\ 
$n_1' = 15$ & 0.083(0.004) & 0.267(0.005) & 0.083(0.004) & \textbf{0.082(0.004)}\\ 
$n_1' = 20$ & 0.068(0.003) & 0.267(0.005) & 0.070(0.003) & \textbf{0.068(0.003)}\\ 
$n_1' = 25$ & 0.061(0.002) & 0.267(0.005) & 0.063(0.003) & \textbf{0.061(0.002)}\\ 
$n_1' = 30$ & 0.055(0.002) & 0.267(0.005) & 0.056(0.002) & \textbf{0.054(0.002)}\\ 
$n_1' = 35$ & 0.051(0.002) & 0.267(0.005) & 0.053(0.002) & \textbf{0.050(0.002)}\\ 
$n_1' = 40$ & 0.046(0.002) & 0.267(0.005) & 0.048(0.002) & \textbf{0.045(0.002)}\\ 
$n_1' = 45$ & 0.044(0.002) & 0.267(0.005) & 0.045(0.002) & \textbf{0.044(0.002)}\\ 
$n_1' = 50$ & 0.042(0.002) & 0.267(0.005) & 0.043(0.002) & \textbf{0.041(0.002)}\\ 
\hline
     \multicolumn{2}{c}{$ \sigma_1 = 0.5, \rho = 0.6$}    \\
\hline
$n_1' = 5$ & 0.211(0.010) & 0.267(0.005) & \textbf{0.153(0.007)} & 0.174(0.009)\\ 
$n_1' = 10$ & 0.110(0.005) & 0.267(0.005) & \textbf{0.086(0.004)} & 0.087(0.004)\\ 
$n_1' = 15$ & 0.081(0.003) & 0.267(0.005) & 0.068(0.003) & \textbf{0.067(0.003)}\\ 
$n_1' = 20$ & 0.066(0.003) & 0.267(0.005) & 0.059(0.002) & \textbf{0.059(0.002)}\\ 
$n_1' = 25$ & 0.059(0.002) & 0.267(0.005) & 0.055(0.002) & \textbf{0.054(0.002)}\\ 
$n_1' = 30$ & 0.053(0.002) & 0.267(0.005) & 0.051(0.002) & \textbf{0.050(0.002)}\\ 
$n_1' = 35$ & 0.049(0.002) & 0.267(0.005) & 0.049(0.002) & \textbf{0.047(0.002)}\\ 
$n_1' = 40$ & 0.045(0.002) & 0.267(0.005) & 0.045(0.002) & \textbf{0.044(0.002)}\\ 
$n_1' = 45$ & 0.043(0.002) & 0.267(0.005) & 0.044(0.002) & \textbf{0.043(0.002)}\\ 
$n_1' = 50$ & 0.041(0.002) & 0.267(0.005) & 0.043(0.002) & \textbf{0.041(0.002)}\\ 
\hline
    \multicolumn{2}{c}{$ \sigma_1 = 0.5, \rho = 0.9$}    \\
\hline
$n_1' = 5$ & 0.212(0.009) & 0.267(0.005) & 0.100(0.005) & \textbf{0.078(0.004)}\\ 
$n_1' = 10$ & 0.116(0.005) & 0.267(0.005) & 0.063(0.003) & \textbf{0.053(0.002)}\\ 
$n_1' = 15$ & 0.082(0.003) & 0.267(0.005) & 0.054(0.002) & \textbf{0.048(0.002)}\\ 
$n_1' = 20$ & 0.064(0.003) & 0.267(0.005) & 0.049(0.002) & \textbf{0.046(0.002)}\\ 
$n_1' = 25$ & 0.056(0.002) & 0.267(0.005) & 0.047(0.002) & \textbf{0.045(0.002)}\\ 
$n_1' = 30$ & 0.051(0.002) & 0.267(0.005) & 0.046(0.002) & \textbf{0.043(0.002)}\\ 
$n_1' = 35$ & 0.046(0.002) & 0.267(0.005) & 0.044(0.002) & \textbf{0.043(0.002)}\\ 
$n_1' = 40$ & 0.044(0.002) & 0.267(0.005) & 0.043(0.002) & \textbf{0.042(0.002)}\\ 
$n_1' = 45$ & 0.042(0.002) & 0.267(0.005) & 0.043(0.002) & \textbf{0.042(0.002)}\\ 
$n_1' = 50$ & 0.041(0.002) & 0.267(0.005) & 0.043(0.002) & \textbf{0.041(0.002)}\\ 
\hline
\hline
    \end{tabular}
    }
     \caption{ {\bf Mean square error~(MSE) of $a_0$ for different procedures when both sample sizes $n_1=n_2=50$.} We present the average MSE based on $1000$ replications, with the number in the parenthesis indicating the SEM. This comparison provides insights into the performance of these estimation procedures under different parameter settings. \label{tab:MSE_S1}}
\end{table}

\begin{figure}
    \centering
    \includegraphics[width=\textwidth]{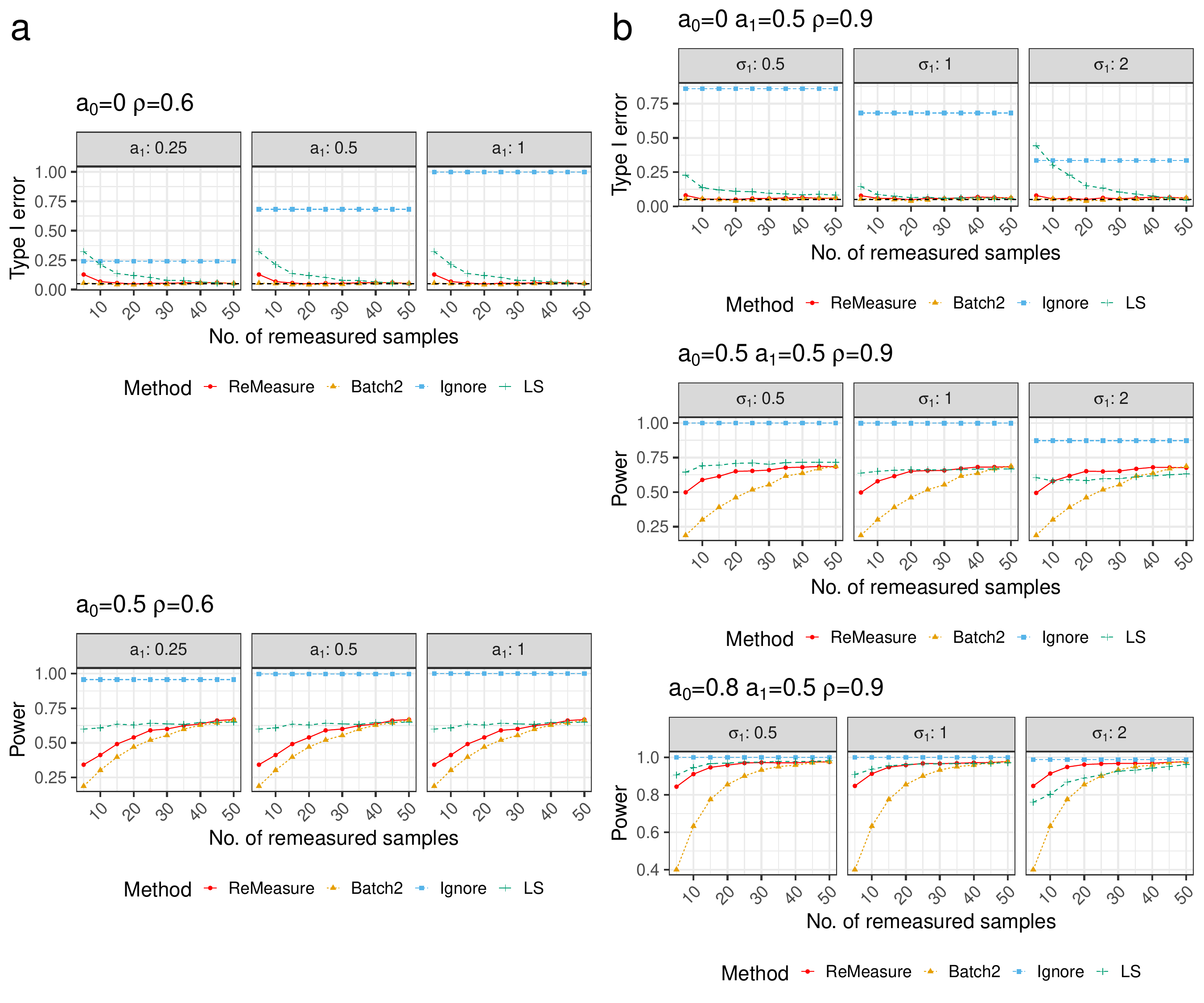}
    \caption{{\bf Assessing empirical type I error and power for testing the biological effect $a_0 = 0$ for different procedures with sample sizes $n_1=n_2=50$.} (a) Panels organized from left to right present the results under different values of $a_1$ (the batch location parameter), based on $1000$ replications. The proposed estimator of $a_0$ does not depend on the batch location. (b) The ``ReMeasure" and ``Batch2" are not affected by the choices of $\sigma_1$. The dashed line indicates the nominal type I error rate used. }
    \label{fig:Power_a1_s1}
\end{figure}

\begin{figure}
    \centering
    \includegraphics[width = \textwidth]{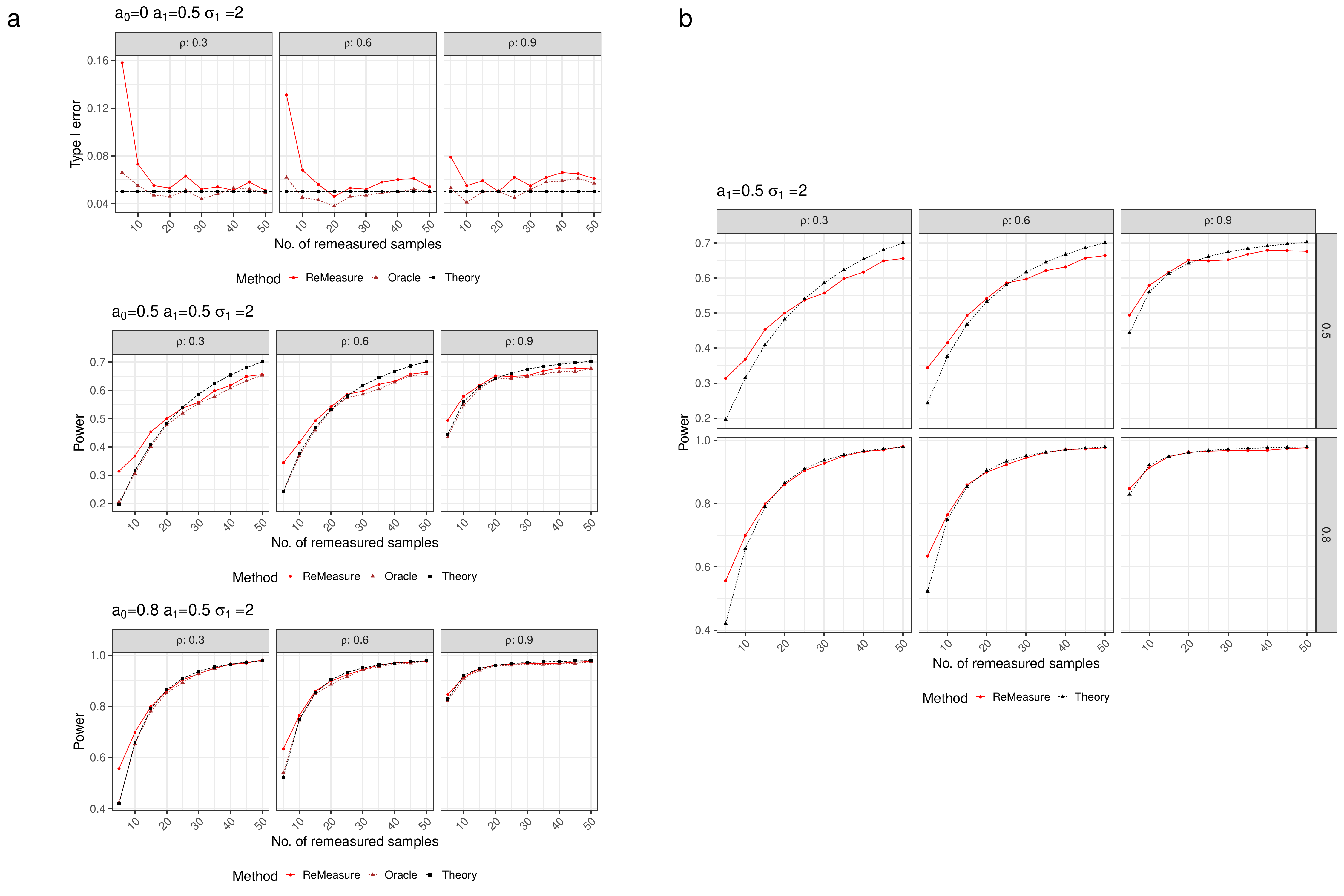}
    \caption{{\bf Comparison of empirical type I error and statistical power for ``ReMeasure", ``Oracle", and ``Theory" with $n_1 = n_2 = 50$.} (a) In the ``Oracle" scenario, we assume $\sigma_1, \sigma_2,$ and $ \rho$ are known. The ``Theory" curve is derived based on the theoretical power formula. (b) A magnified view of the power comparison plot between ``ReMeasure" and ``Theory," offering a clearer sight.}
    \label{fig:Theory_Oracle}
\end{figure}

\begin{figure}[ht!]
    \centering
    \includegraphics[width = 0.9\textwidth]{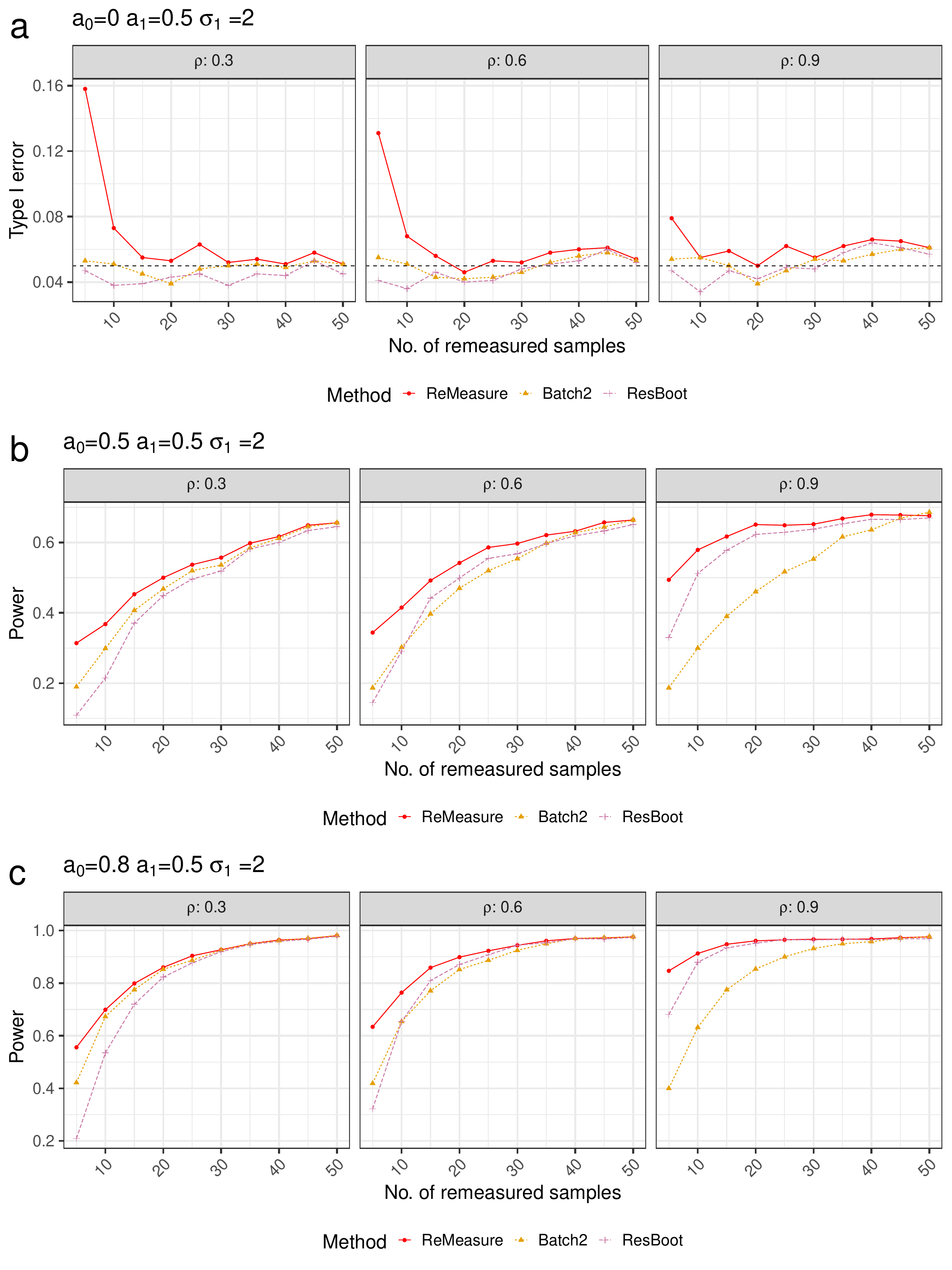}
    \caption{ {\bf 
    Assessing empirical type I error and power of the residual bootstrap across  parameter settings with sample sizes $n_1=n_2=50$. } The bootstrap method can control the type I error at small remeasured sample sizes and deliver comparable power as ``ReMeasure" when the remeasured sample size is large.  The between-batch correlation, $\rho$, takes values of  $0.3, 0.6 $, and $0.9$ from the left to the right panel.  }
    \label{fig:Power_S1_Boot}
\end{figure}

\begin{figure}
    \centering
    \includegraphics[width=\textwidth]{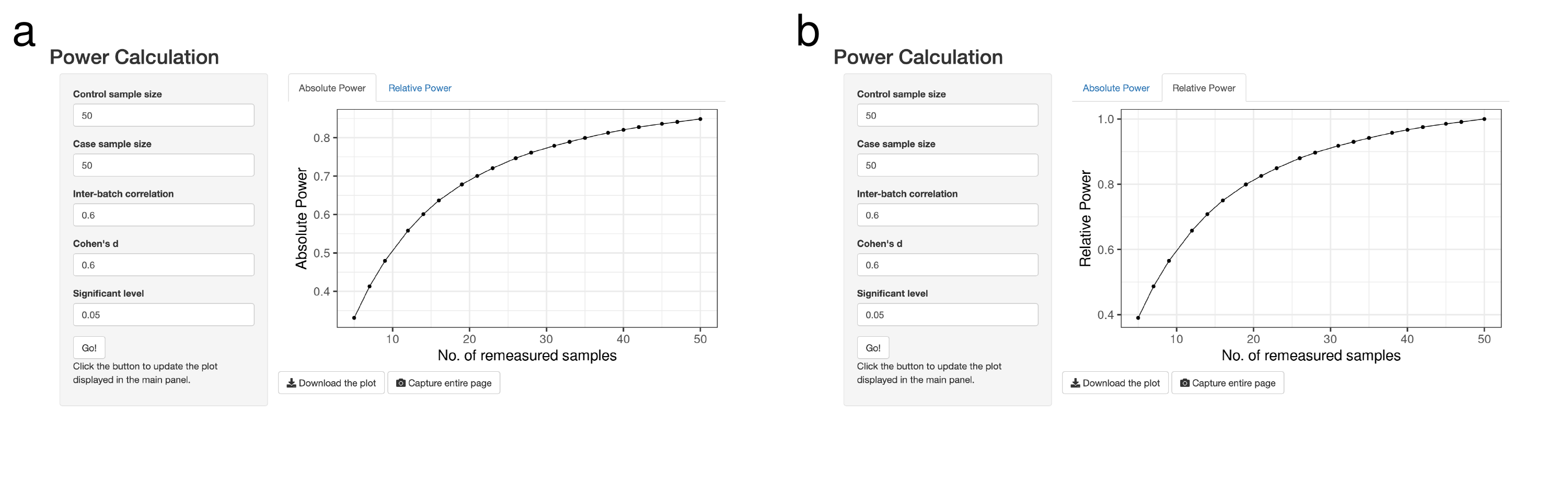}
    \caption{{\bf An example of power analysis using the Shiny app for a confounded case-control study with sample measurement.} (a) The absolute power vs. No. of remeasured samples. (b) The relative power vs. No. of remeasured samples.  ~\label{fig:absolute&ratio}}
\end{figure}

\begin{figure}
    \centering
    \includegraphics[width=\textwidth]{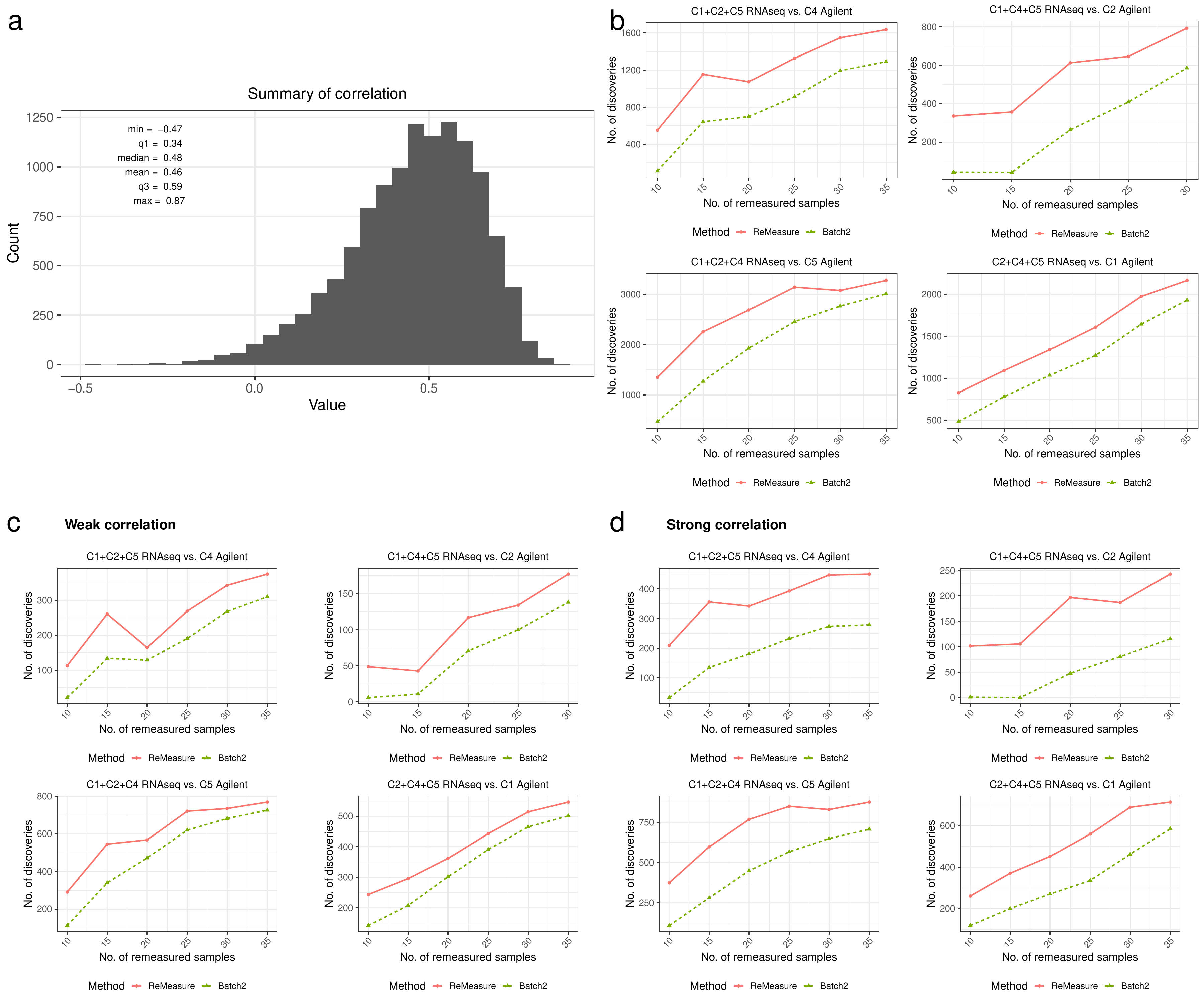}
    \caption{{\bf Correlation of the gene expression between platforms and significant gene discovery in ovarian cancer dataset.} (a) Distribution of the correlation coefficients of the gene expression level between the Agilent and RNA-Seq platform estimated based on 47 common samples. (b), (c) and (d) depicts the number of discovered significant genes vs. the number of remeasured samples, comparing ``ReMeasure" to ``Batch2".   (b) All genes are used. (c) A quarter of the genes with the lowest correlation are used. (d) A quarter of the genes with the highest correlation are used. In the title, ``C1+C2+C5 RNAseq vs. C4 Agilent"  refers to comparing combined ``C1-MES," ``C2-IMM," and ``C5-PRO" subtypes from the RNAseq platform to the ``C4-DIF" subtype from the Agilent platform. The same explanation applies to other titles.}
    \label{fig:Power_C2C4C5_RNAseq}
\end{figure}

\begin{figure}[ht!]
    \centering
    \includegraphics[width = 0.9\textwidth]{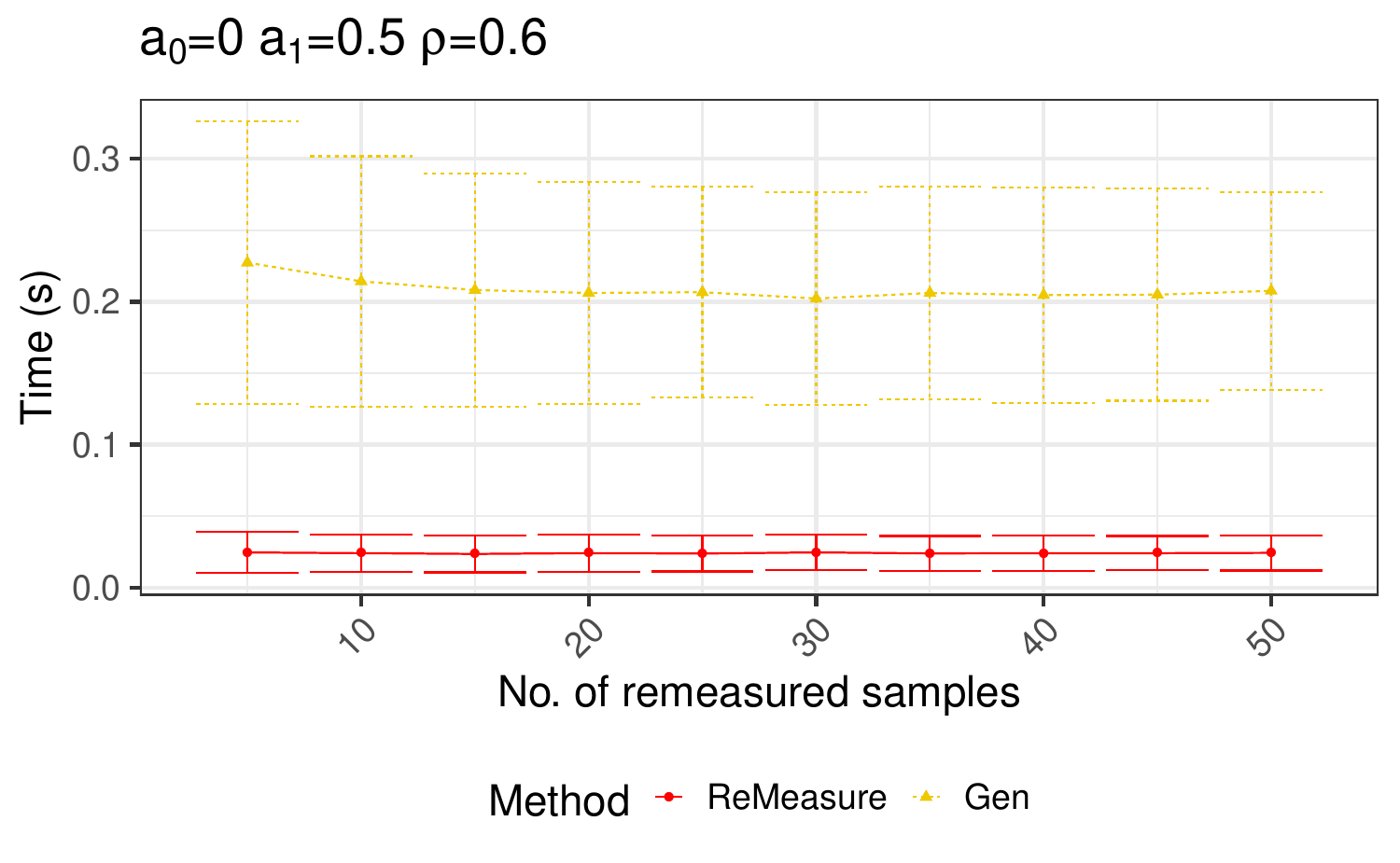}
    \caption{ \textbf{Computational time of ``ReMeasure" compared to the generic MLE optimization approach (``Gen").}  ``Gen" uses the \textit{optim} function in the R \textit{stat} package. ``ReMeasure" is $>10$ times faster than ``Gen" ($n_1 = n_2 = 50$, and $500$ replications are conducted). Data are presented as mean values +/- SD. We run all methods on the same computation platform with $2.40$ GHz Intel (R) Xeon (R) E5-2680 v4 28-Core CPU. }
    \label{fig:Time_S1}
\end{figure} 

\subsection{The effect of the batch location parameter}
To examine the influence of batch location effect $a_1$ on method performance, we fix the batch scale parameter $\sigma_1 = 0.5$ and the effect size $a_0 = 0.5$. Supplementary Figure~\ref{fig:Power_a1_s1}a shows that $a_1$ has substantial impacts on ``Ignore'' while the other methods are not affected by $a_1$. 

\subsection{Performance under large sample sizes}

\begin{figure}
    \centering
    \includegraphics[width=\textwidth]{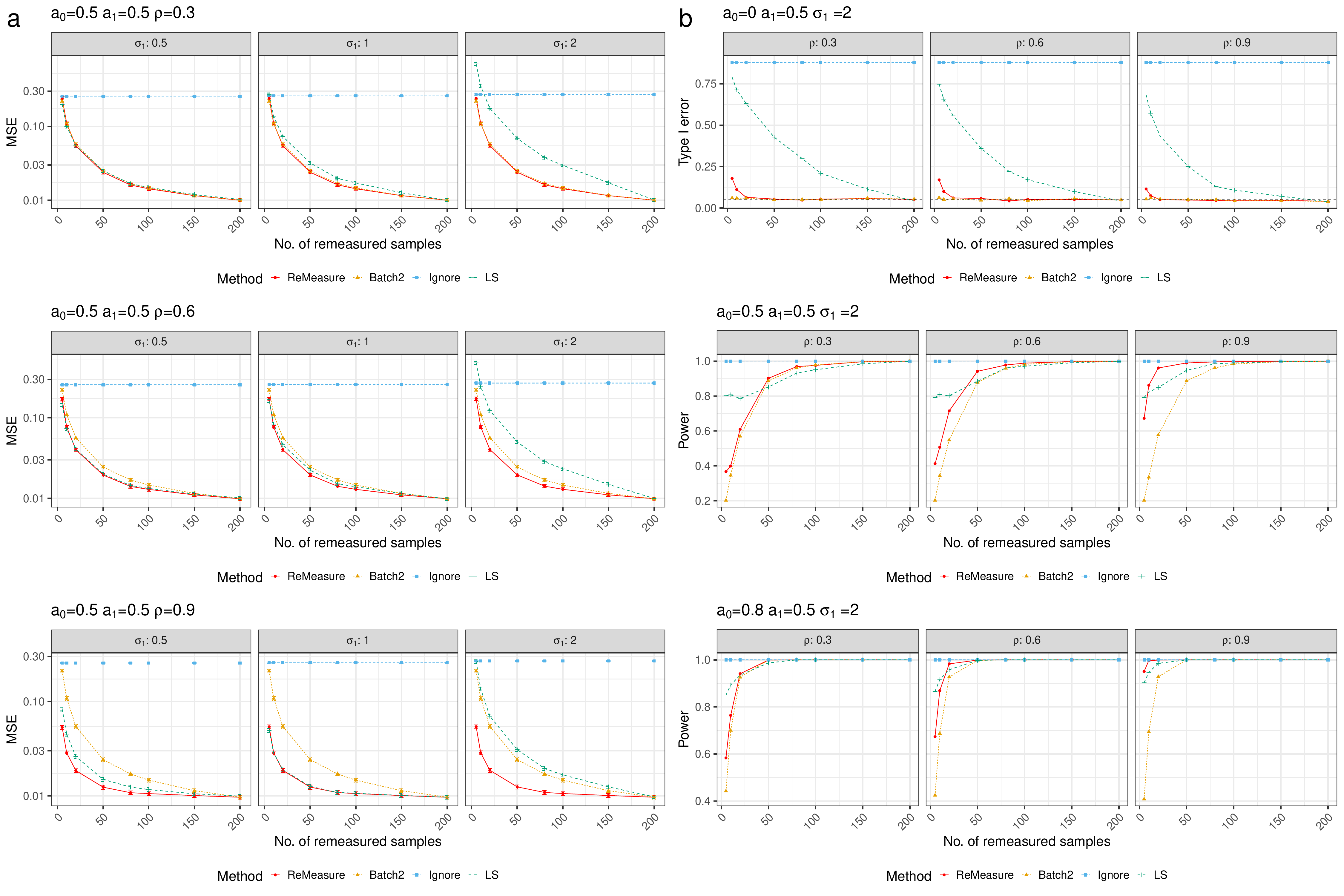}
    \caption{{\bf Mean square errors and statistical power of different procedures with sample sizes $n_1=n_2=200$.} (a) Data are presented as mean values +/- SEM at varying noise levels ($\sigma_1$).  y-axis is presented in $\log_{10}$ scale. (b) The type I error and power at different between-batch correlations ($\rho$). The dashed line indicates the $5\%$ nominal type I error rate used. All results are based on $1000$ replications.}
    \label{fig:200}
\end{figure}
In the case of large sample sizes (i.e., $n_1 = n_2 = 200$), we present the MSE and power curves in Supplementary Figures~\ref{fig:200}a and \ref{fig:200}b.

\subsection{Performance under non-Gaussian noises}
\begin{figure}
    \centering
    \includegraphics[width=\textwidth]{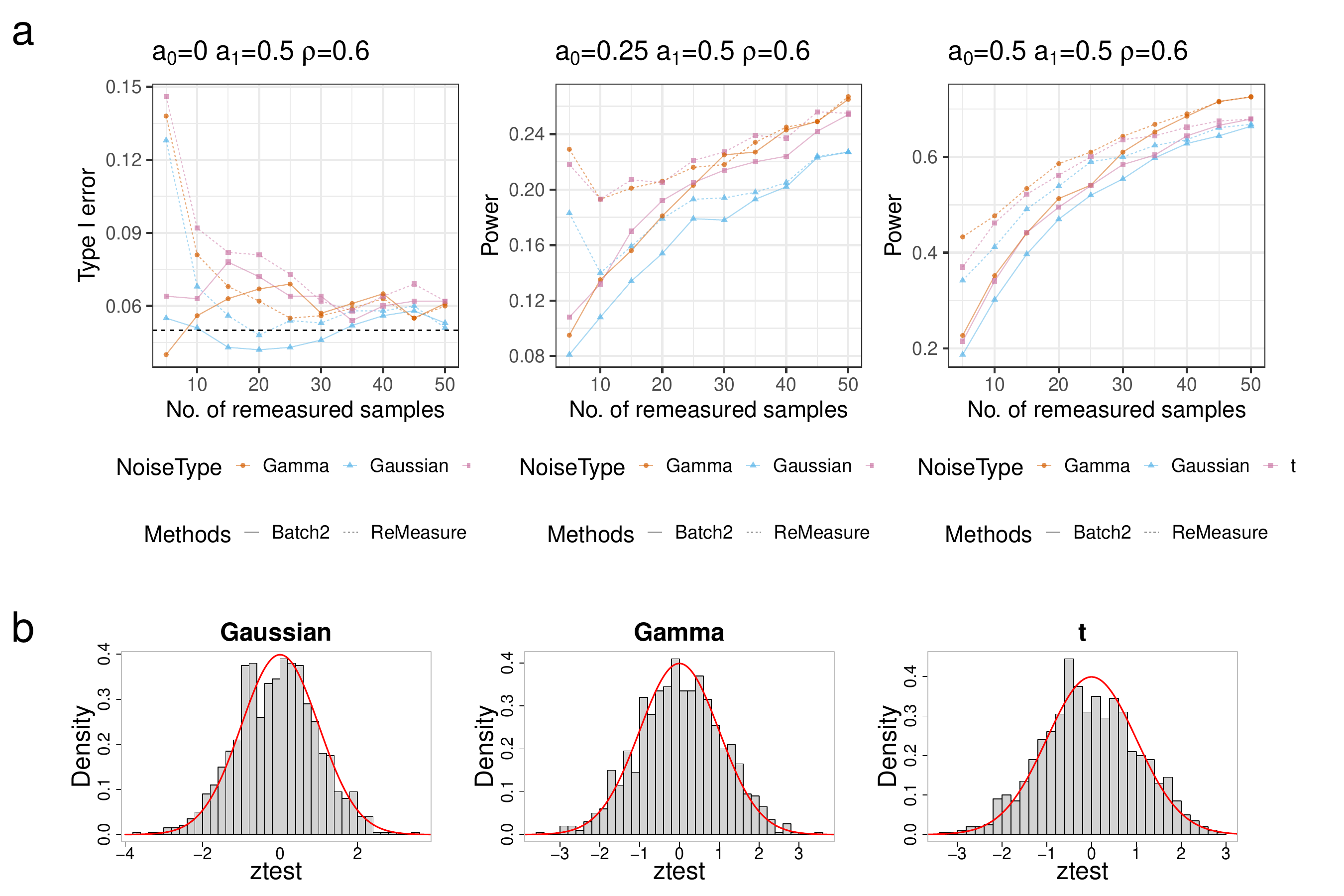}
    \caption{{\bf Impact of noise distributions on type I error, power, and Z-statistics.} (a) Type I error and power curves under different noise distributions: centered gamma, Gaussian, and $t_6$ distributions. (b) The histograms of the Z-statistics are calculated using the proposed method when $n_1' = 50$ under three different noise distributions: centered gamma, Gaussian, and $t_6$.}
    \label{fig:NG}
\end{figure}
For the proposed method to work, the error in the regression model does not have to follow the Gaussian distribution as we stated in theory.  Here we consider the cases where errors follow the centered gamma distribution with the shape parameter $2$ and scale parameter $1$. We also consider the student t-distribution with degrees of freedom equal to $6$.  Supplementary Figure~\ref{fig:NG}a presents the power curves under these noise distributions. The power behaviors in these three cases have similar patterns. The power is slightly higher under the non-Gaussian error at the price of a more inflated type I error compared to the Gaussian case.

The histograms of the z-statistics in Supplementary Figure~\ref{fig:NG}b are close to that of the standard normal distribution under different noise distributions, which empirically justifies the asymptotic normal approximation.

\begin{figure}
    \centering
    \includegraphics[width = \textwidth]{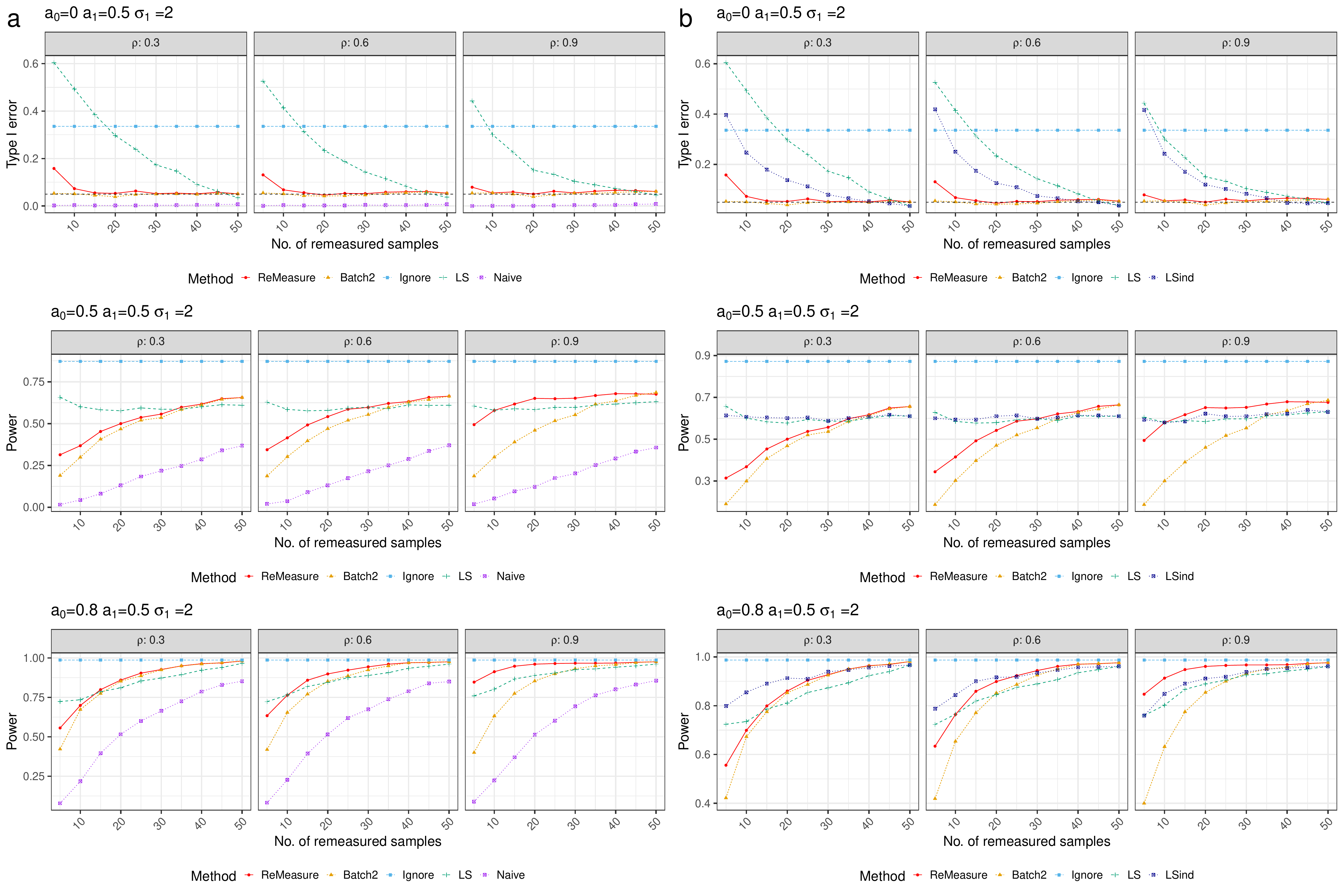}
    \caption{{\bf The empirical type I error and power for testing the biological effect $a_0 = 0$ under different parameter settings.} The dashed line indicates the nominal type I error rate used. (a) ``naive" is the method that fits a linear regression model based on all samples adjusting the batch variable and ignoring the repeated measurement. (b) ``LSind" refers to the location-scale matching method using all control samples in $C_1$ and $C_2$.}
    \label{fig:naive_LSind}
\end{figure}
\subsection{Comparison to the naive least square approach}
We compare to the naive approach based on the model $Y \sim X + \textrm{Batch} + Z$. This approach neglects the repeated measure nature and the heterogeneity of variances, which may lead to a reduction in statistical power. Supplementary Figure~\ref{fig:naive_LSind}a shows that its power is substantially lower than other competing methods on the same simulated datasets.

\subsection{Comparison to the location-scale matching method using all the control samples}

 We compare to the location-scale matching method using all independent control samples in the first batch ($C_1$, $C_2$). The new approach, ``LSind", has milder type I error inflation than the original ``LS" based only on the controls that are remeasured. However, its performance deteriorates in the small-sample setting, with a substantially larger type I error above the nominal level compared to the  ``ReMeasure" and ``Batch2" methods (Supplementary Figure \ref{fig:naive_LSind}b).




\end{appendices}

\end{document}